\documentclass[twocolumn]{aastex631}
\usepackage{textcomp}

\graphicspath{{./}{Final_figures/}}

\newcommand{\MSun} {\mbox{$M_{\odot}$}}

\newcommand{\be}{\begin{eqnarray}}
\newcommand{\ee}{\end{eqnarray}}

\begin{document}

\title{Trans-Neptunian Object dynamics even better explained by a stellar flyby after 4.5 Gyr of evolution}

\author[0000-0002-5003-4714]{Susanne Pfalzner}
\affiliation{J\"ulich Supercomputing Center, Forschungszentrum J\"ulich, 52428 J\"ulich, Germany}

\author[0009-0001-5904-1742]{Frank W. Wagner}
\affiliation{J\"ulich Supercomputing Center, Forschungszentrum J\"ulich, 52428 J\"ulich, Germany}

\author[0009-0000-7124-5329]{Marco Bischoff}
\affiliation{J\"ulich Supercomputing Center, Forschungszentrum J\"ulich, 52428 J\"ulich, Germany}

\begin{abstract}
The Trans-Neptunian objects (TNOs) formed together with the planets from a flat disc of gas and dust but now orbit the Sun mostly on inclined, eccentric orbits. One explanation for the TNOs' orbits is a close flyby of another star. One with a mass of $M_p =$~0.8~\MSun\ at $q_p =$~110~au and $i_p =$~70° fits the observations. However, such close encounters were more frequent when the Sun was young and still part of its birth cluster. Assuming that the flyby happened then, we use numerical $N$-body simulations to model how the TNOs' orbits changed due to interactions with the planets over the 4.56~Gyr since the Sun formed. We find that the cold Kuiper belt region lost about 80\% and the hot Kuiper belt population about 40\% of its initial population. About 7\% -- 8\% of the TNOs were injected into the planet region, but almost all (99\%) were ejected afterwards. By contrast, the orbital parameters of distant Sedna-like objects ($q >$~60~au) remained basically unchanged. Surprisingly, in sum, these changes improve the fit to the observed TNO population even more, strengthening the argument for a close flyby to the Solar System. The remaining differences concern the location of the cold population and the inclination distribution of the Sedna population. We discuss possible reasons and steps to resolve these issues.
\end{abstract}


\section{Introduction}
\label{sec:intro}

The orbits of the major planets in the Solar System are nearly circular and coplanar, with low eccentricities and inclinations. In contrast, most minor bodies residing beyond Neptune, known as trans-Neptunian objects (TNOs), exhibit markedly higher eccentricities and inclinations \citep{Kavalaars:2020, Gladman:2021}. These TNOs formed together with the planets from a disc of dust and gas and are remnants of the planetesimal population of the early Solar System. Despite this common formation environment, the dynamical pathways that transformed TNO orbits from primordial near-circularity to their present diversity remain a central question in planetary science.

Traditional explanations assume that TNOs were initially positioned much closer to the Sun and scattered onto their eccentric, inclined orbits through gravitational interactions with the nascent giant planets \citep{Fernandez:1984,Hahn:1999,Gomez:2003,Levison:2008,Raymond:2018}. However, this model cannot on its own account for TNOs with orbits far beyond the gravitational influence of the planets, such as Sedna \citep{Brown:2004}, 2012 VP$_{113}$, 541132 Leleākūhonua, 2023 KQ$_{14}$ \citep{Trujillo:2014,Shephard:2016}, or those exhibiting very high inclinations ($i >$~90°) \citep{Batygin:2016}. An external force could account for these orbits. Such an external force could be provided by a close flyby of another star \citep{Kobayashi:2001,Kenyon:2004,Jilkova:2015,Pfalzner:2018,Moore:2020,Pfalzner:2024a}. Further suggestions to solve this problem include a ninth planet \citep{Batygin:2016,Siraj:2025}, a past flyby of a planet \citep{Huang:2022,Brown:2025}, self-gravitational modulation of the scattered disc \citep{Madigan:2018}, or modified Newtonian dynamics \citep{Brown:2023}.

Here, we focus on the flyby of another star, in particular the specific flyby proposed by \citet{Pfalzner:2024a}, which can quantitatively reproduce many of the observed orbital characteristics, including Sedna-like objects and retrograde TNOs. In this flyby, a star of mass $M_p =$~0.8~\MSun\ passes at a perihelion distance of $q_p =$~110~au on a parabolic orbit inclined by $i_p =$~70°. Thus, a single cause, the flyby of another star, produces most TNO populations in a single event.

However, the frequency of close flybys to the Sun would have been highest when the Sun was still young ($<$~50~Myr) and was still part of or had just left its nascent star cluster. We investigate how the interactions with the giant planets, particularly Neptune, change the orbital properties of the TNOs long after the flyby. The central aim is to determine the effect on different dynamic groups, namely the cold and hot Kuiper belt populations, the scattered disc objects, and the detached objects. During the long-term evolution, a considerable number of TNOs on orbits resonant with Neptune are created. While a future paper will detail these resonant TNOs, this paper focuses on the non-resonant population.

\section{Method}
\label{sec:method}

We modelled the long-term evolution of the TNO population after the close flyby of a star to the Solar System. \citet{Pfalzner:2024a} showed that a perturber star ($M_p =$~0.8~\MSun\ on a parabolic orbit with periastron distance $q_p =$~110~au, inclination $i_p =$~70°, and argument of the periastron $\omega_p =$~80°) effectively reproduces all known dynamic groups among the TNOs \citep{Pfalzner:2024a}. We re-modelled the disc's response to the flyby using the REBOUND code \citep{Rein:2014} with the accurate IAS15 integrator. We treated flyby dynamics as $N$ gravitational three-body interactions between the Sun, the perturber star, and each test particle. We neglect the self-gravity and viscosity effects. Both are relatively small, as the debris disc is essentially gas-free and the mass of the disc is considerably smaller than the Sun's \citep[e.g., ][]{Kobayashi:2001, Musielek:2014}.

\begin{figure}
\centering
\includegraphics[width=\linewidth]{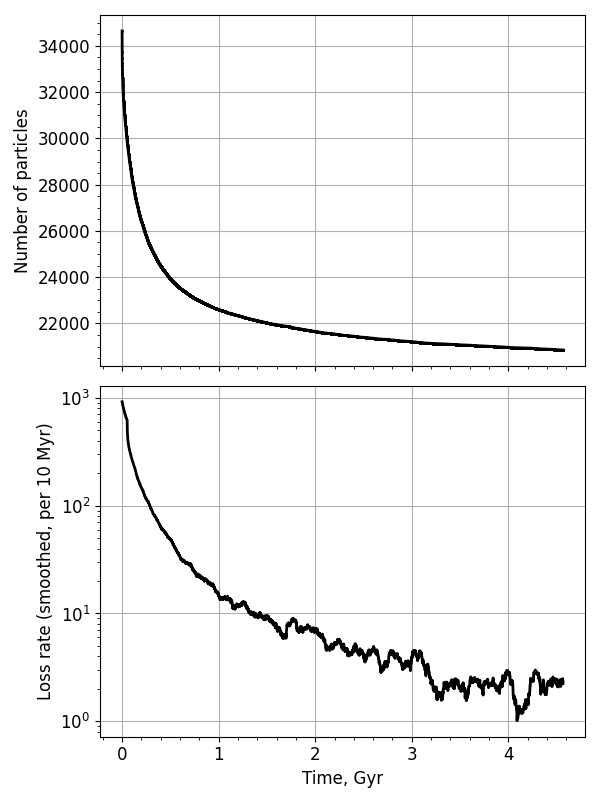}
\caption{Bound TNO population as a function of time elapsed since the stellar flyby. Top: total particle count; bottom: particle loss rate per 10~Myr, smoothed over 100~Myr.}
\label{fig:particle_losses}
\end{figure}

\begin{figure}
\centering
\includegraphics[width=\linewidth]{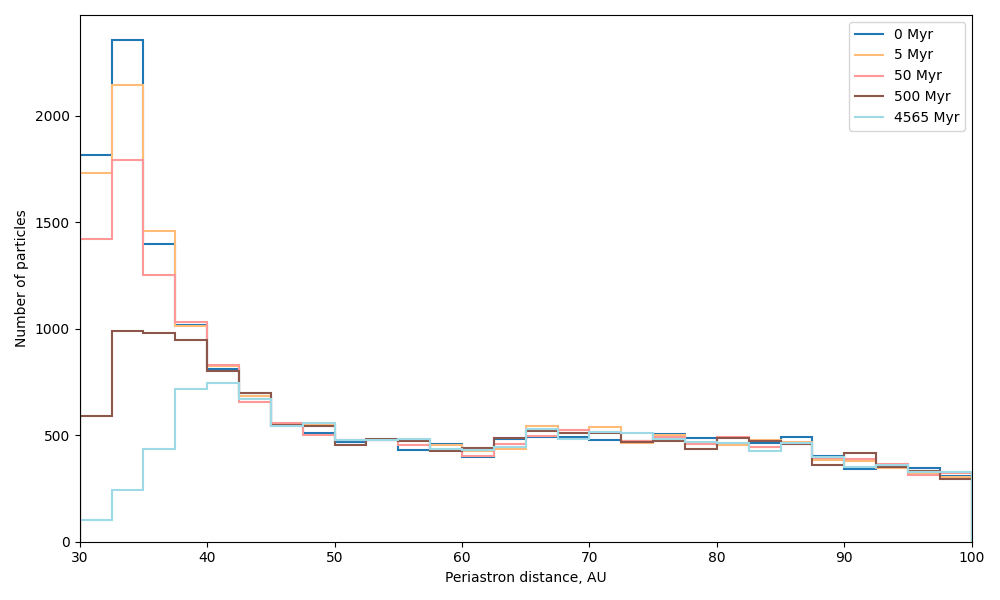}
\caption{Distribution of bound TNOs by periastron distance at several time instances.}
\label{fig:particle_count}
\end{figure}

\begin{figure*}[hp]

\setlength{\tabcolsep}{4pt}
\begin{flushleft}
\begin{tabular}{@{}c c c@{}}
    \begin{minipage}[b]{0.32\textwidth}
    \centering
    \includegraphics[width=\textwidth]{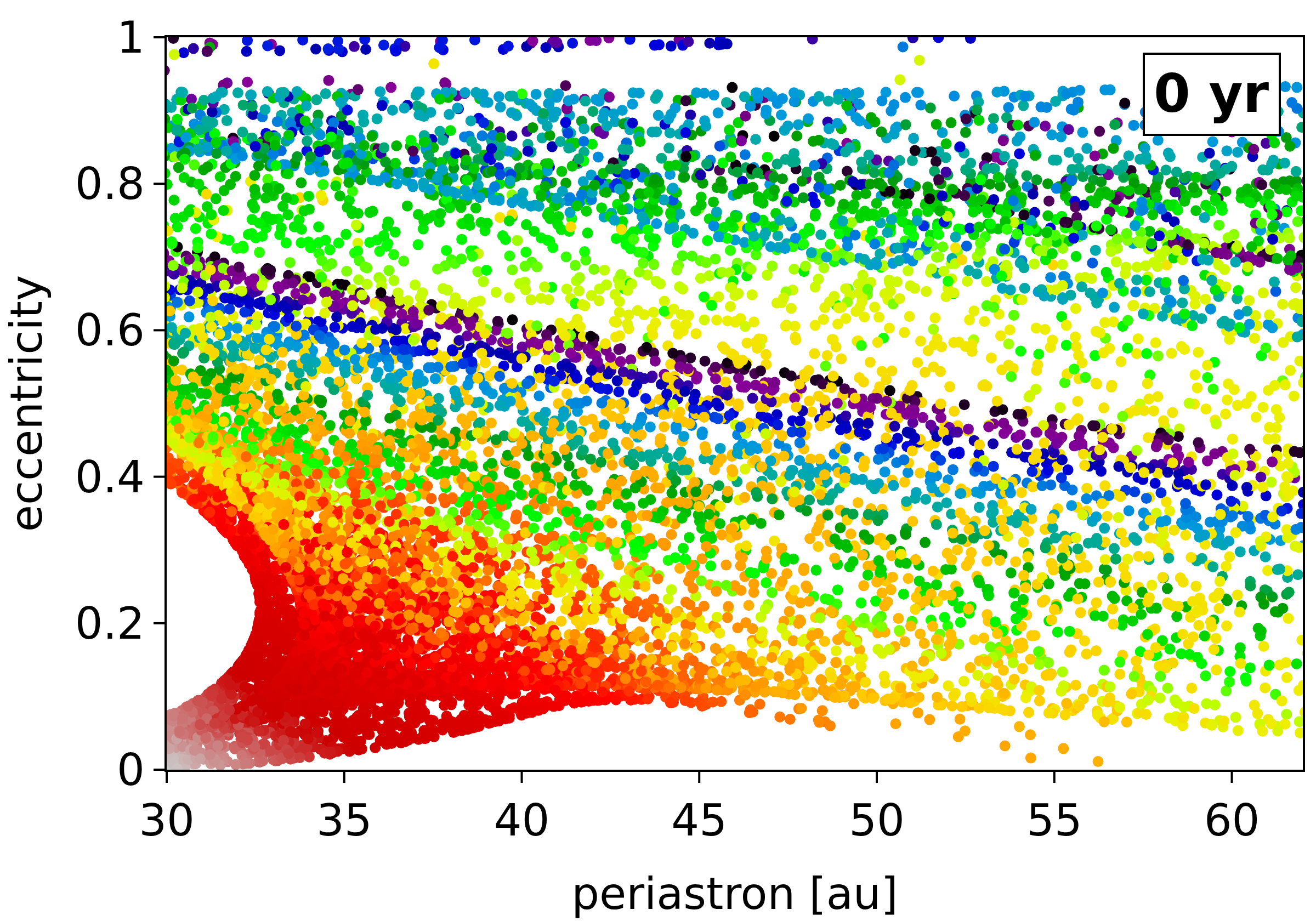}
    \textbf{(a)}
    \end{minipage}
    &
    \begin{minipage}[b]{0.32\textwidth}
    \centering
    \includegraphics[width=\textwidth]{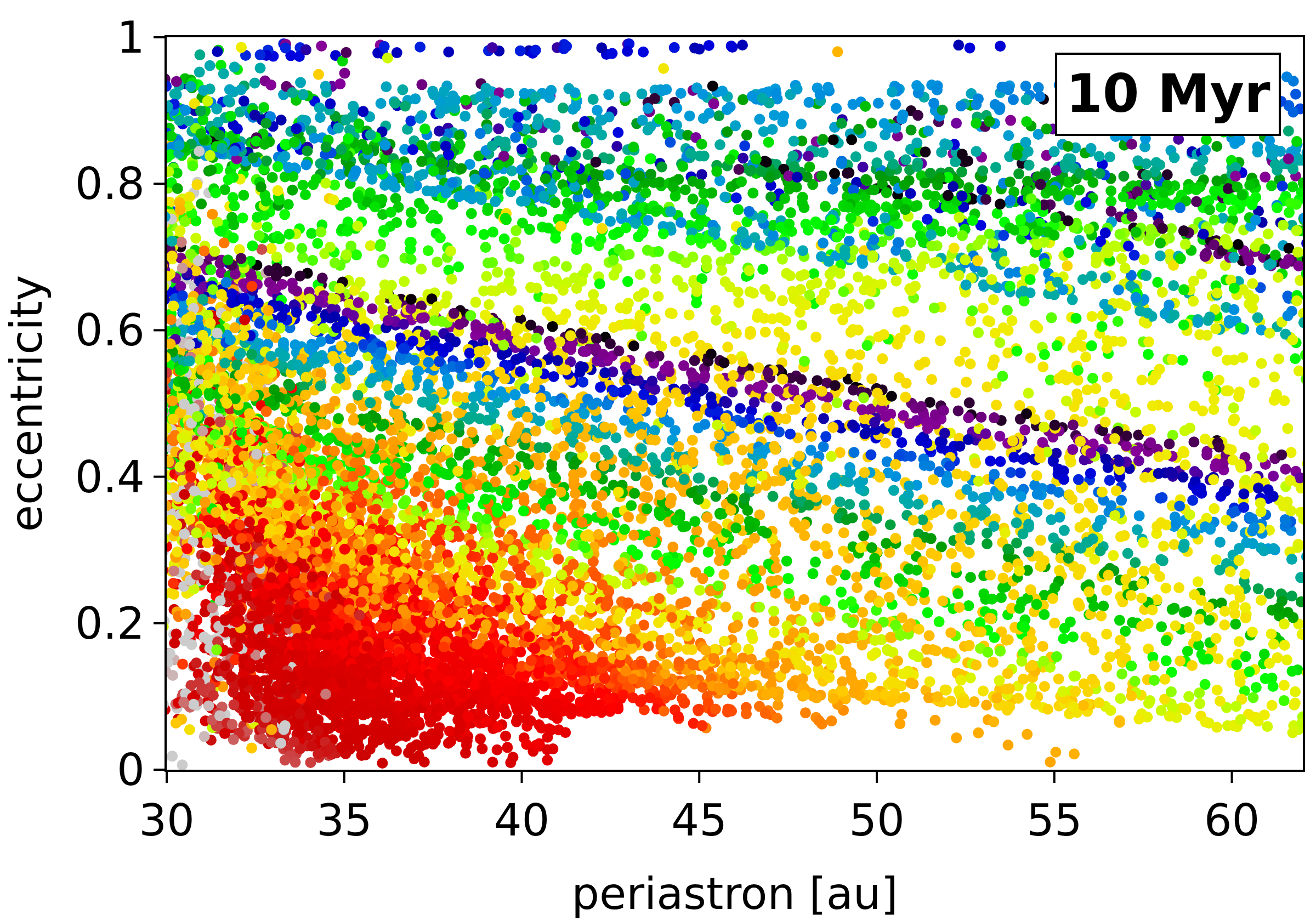}
    \textbf{(b)}
    \end{minipage}
    &
    \begin{minipage}[b]{0.32\textwidth}
    \centering
    \includegraphics[width=\textwidth]{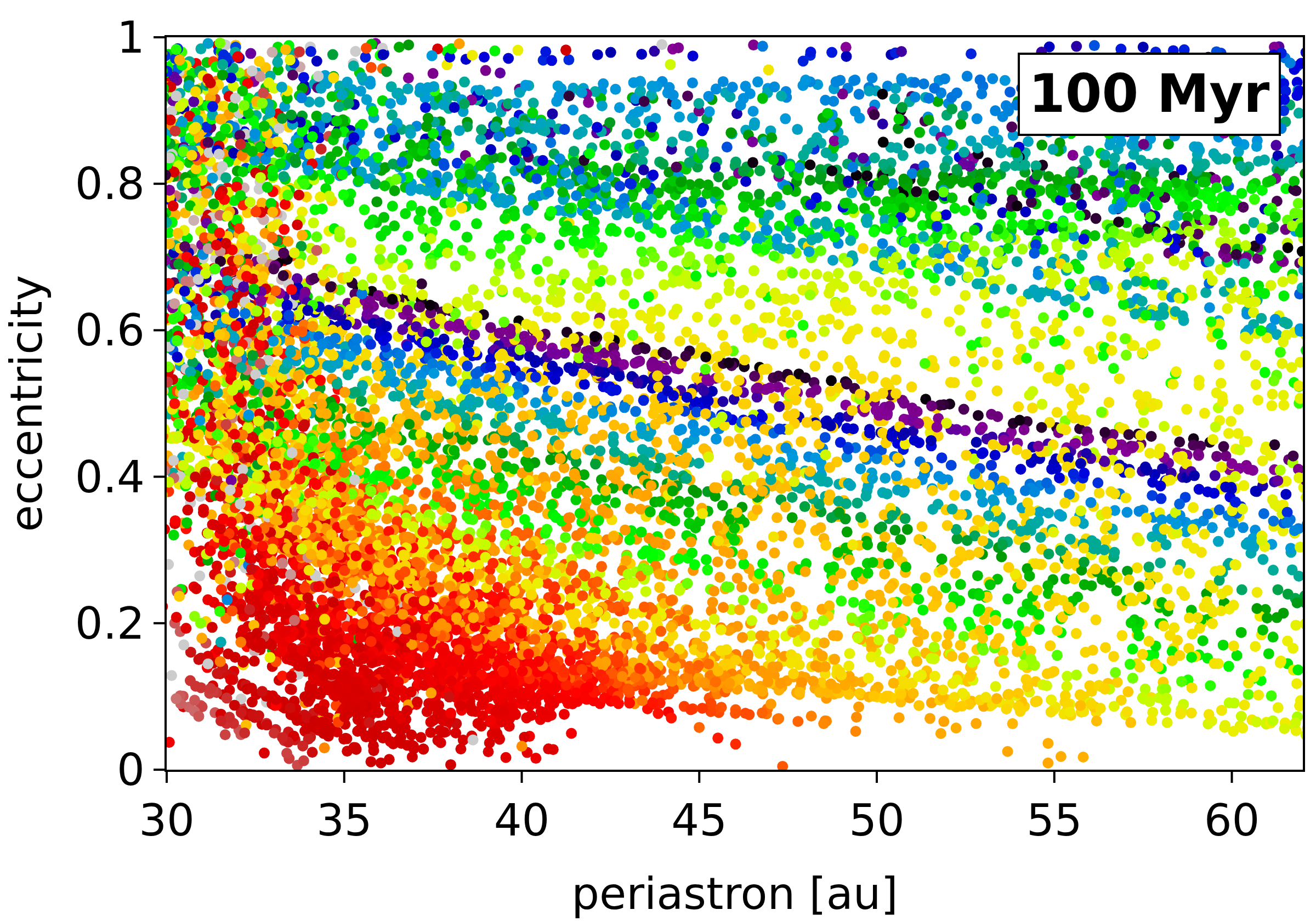}
    \textbf{(c)}
    \end{minipage}
    \\
    \begin{minipage}[b]{0.32\textwidth}
    \centering
    \includegraphics[width=\textwidth]{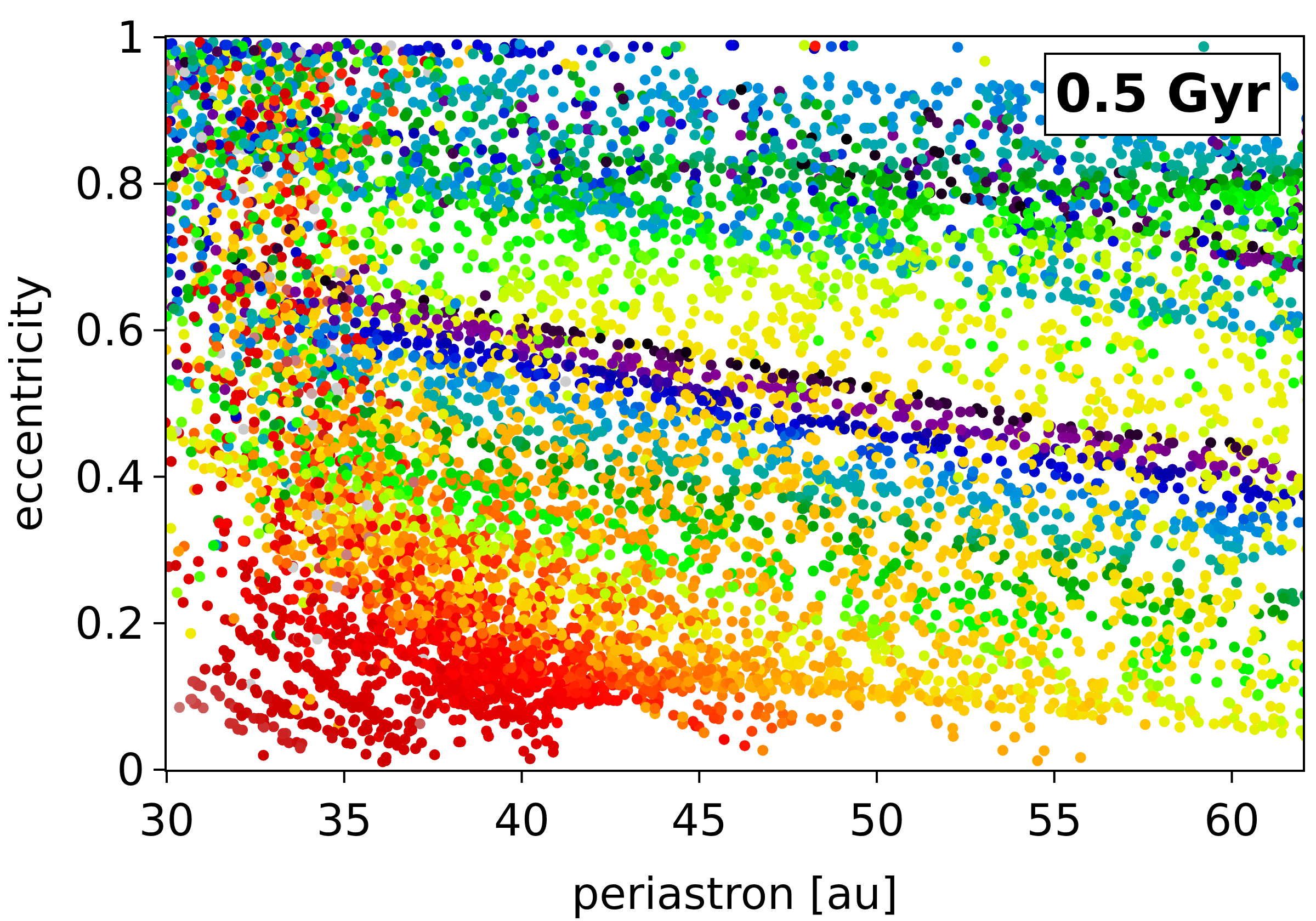}
    \textbf{(d)}
    \end{minipage}
    &
    \begin{minipage}[b]{0.32\textwidth}
    \centering
    \includegraphics[width=\textwidth]{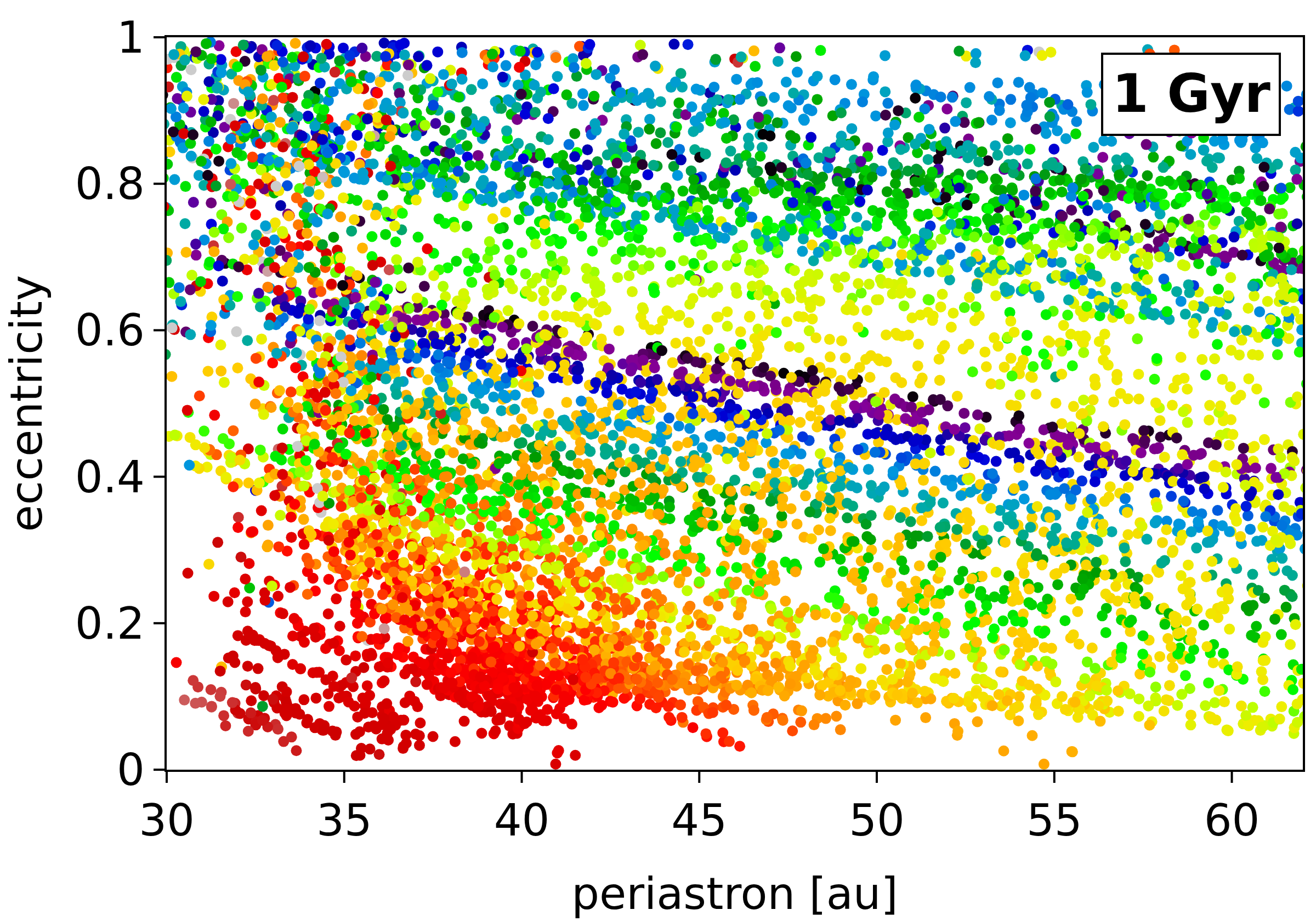}
    \textbf{(e)}
    \end{minipage}
    &
    \begin{minipage}[b]{0.32\textwidth}
    \centering
    \includegraphics[width=\textwidth]{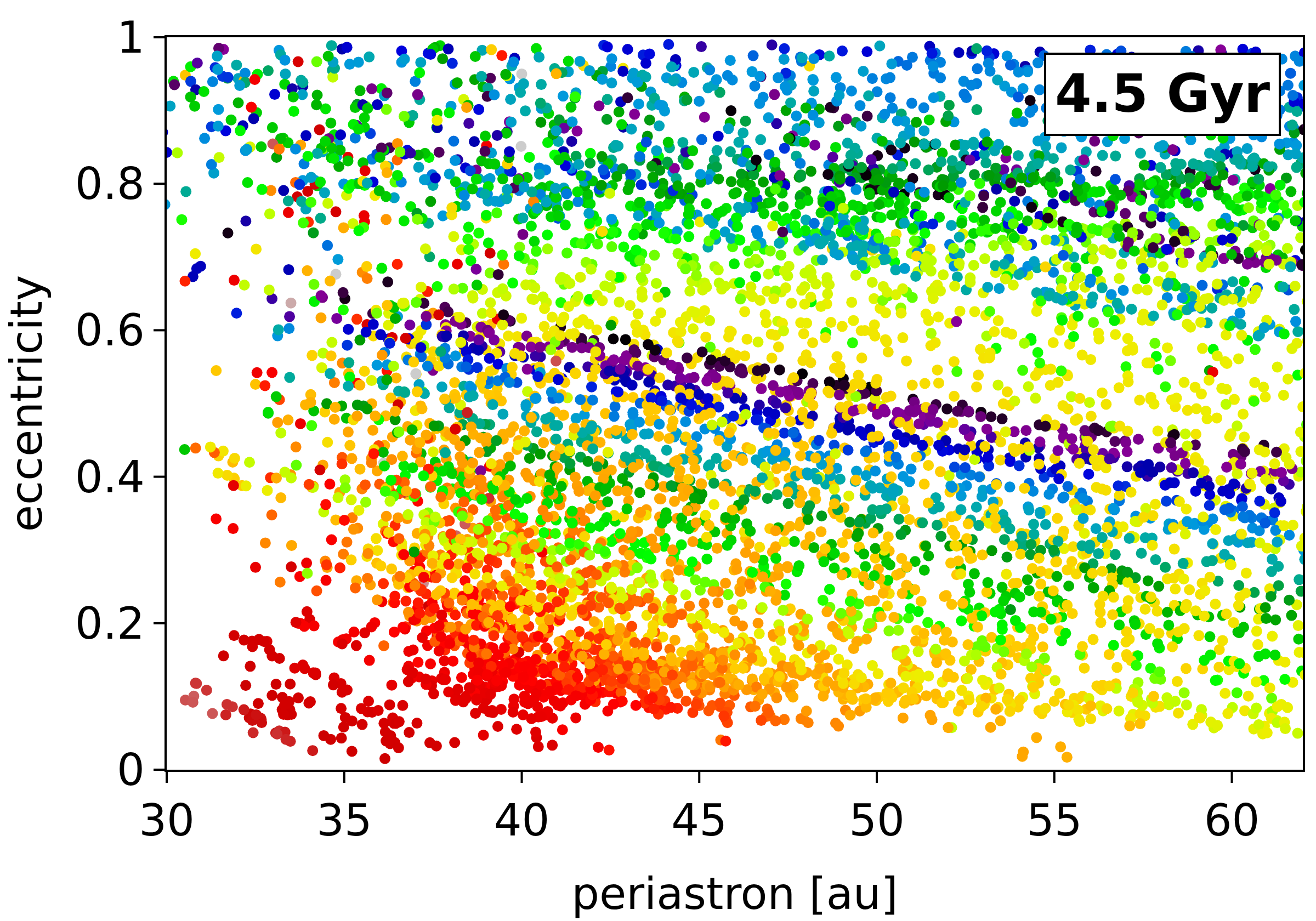}
    \textbf{(f)}
    \end{minipage}
    \\[1cm]
    \begin{minipage}[b]{0.32\textwidth}
    \centering
    \includegraphics[width=\textwidth]{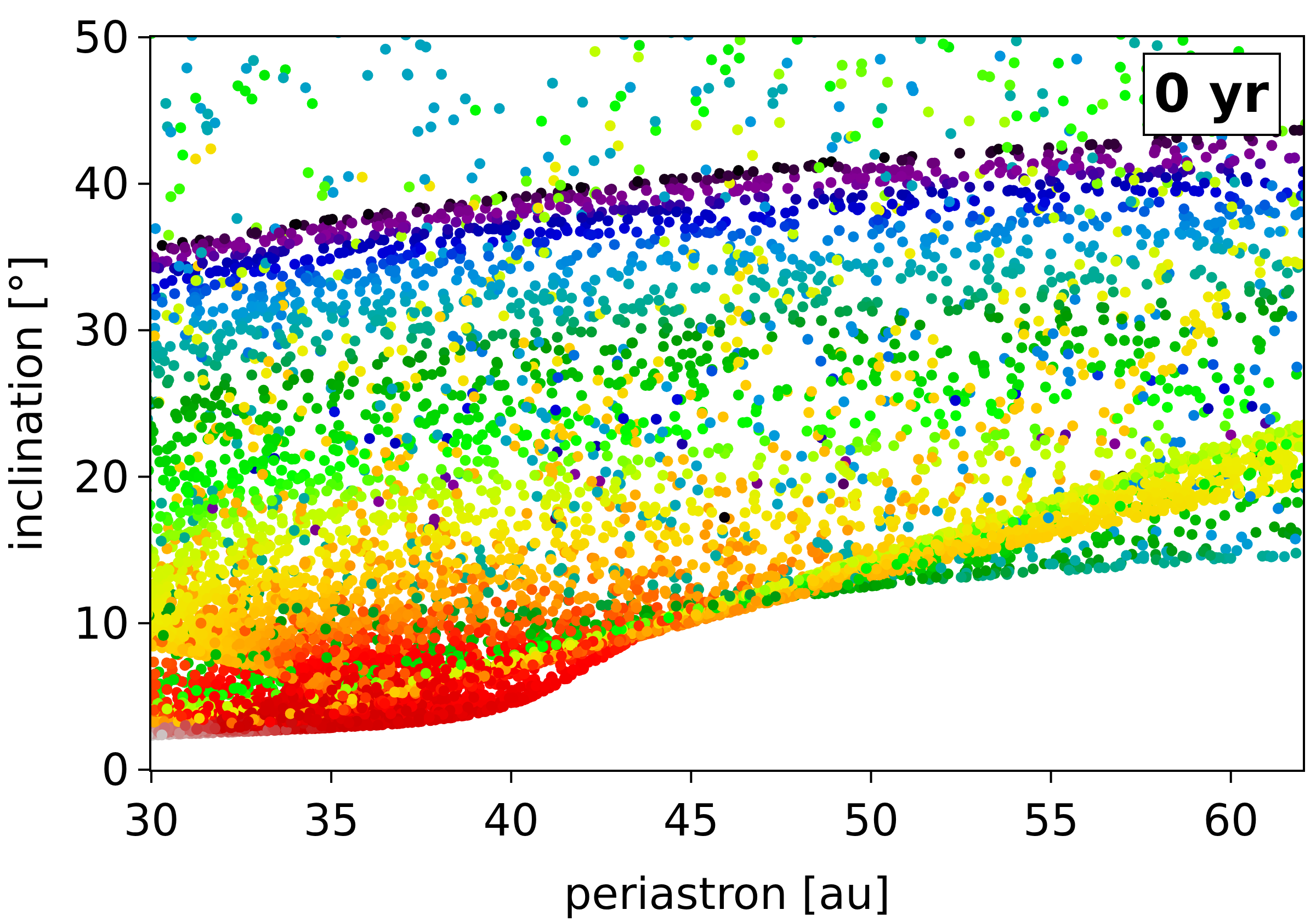}
    \textbf{(g)}
    \end{minipage}
    &
    \begin{minipage}[b]{0.32\textwidth}
    \centering
    \includegraphics[width=\textwidth]{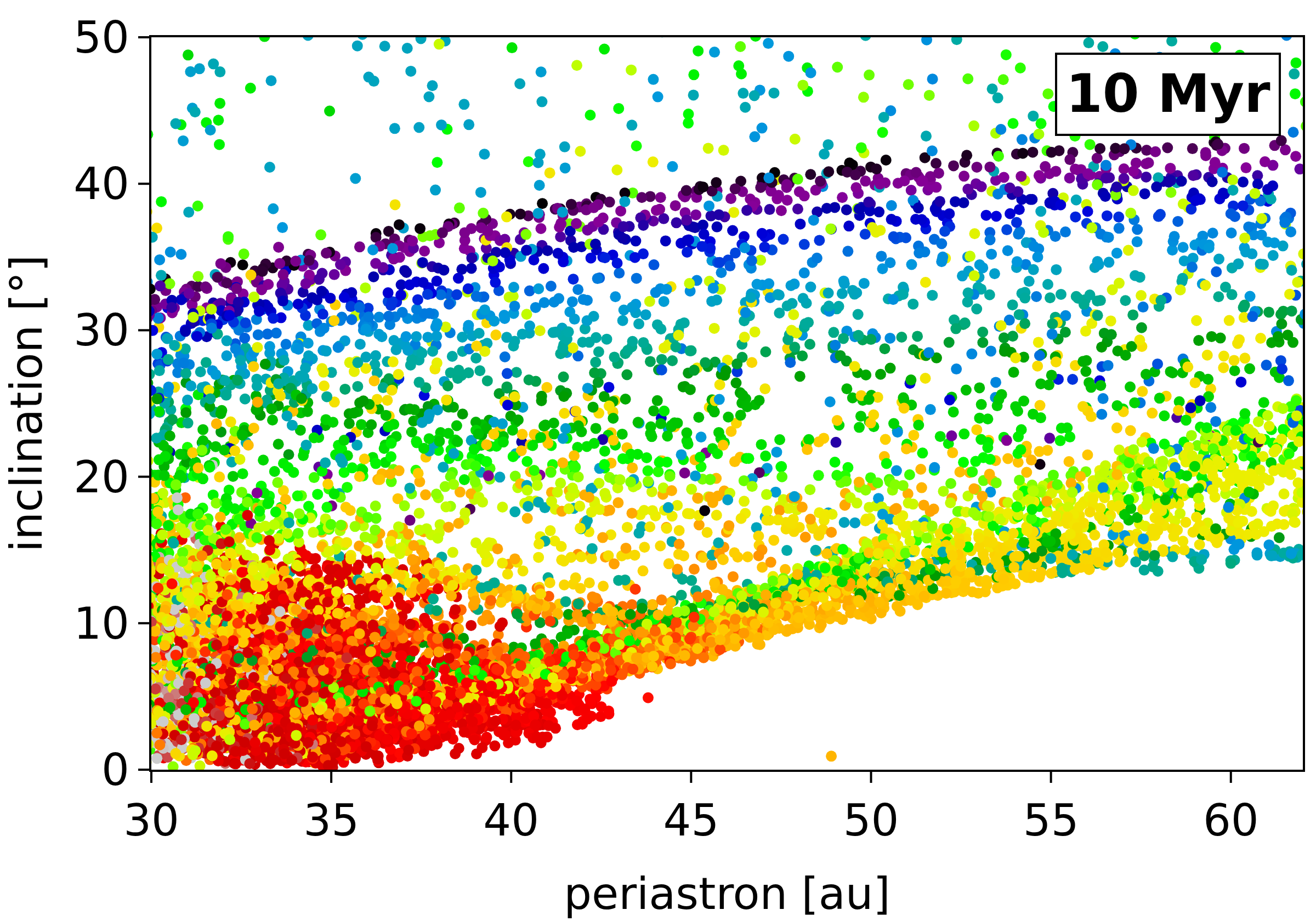}
    \textbf{(h)}
    \end{minipage}
    &
    \begin{minipage}[b]{0.32\textwidth}
    \centering
    \includegraphics[width=\textwidth]{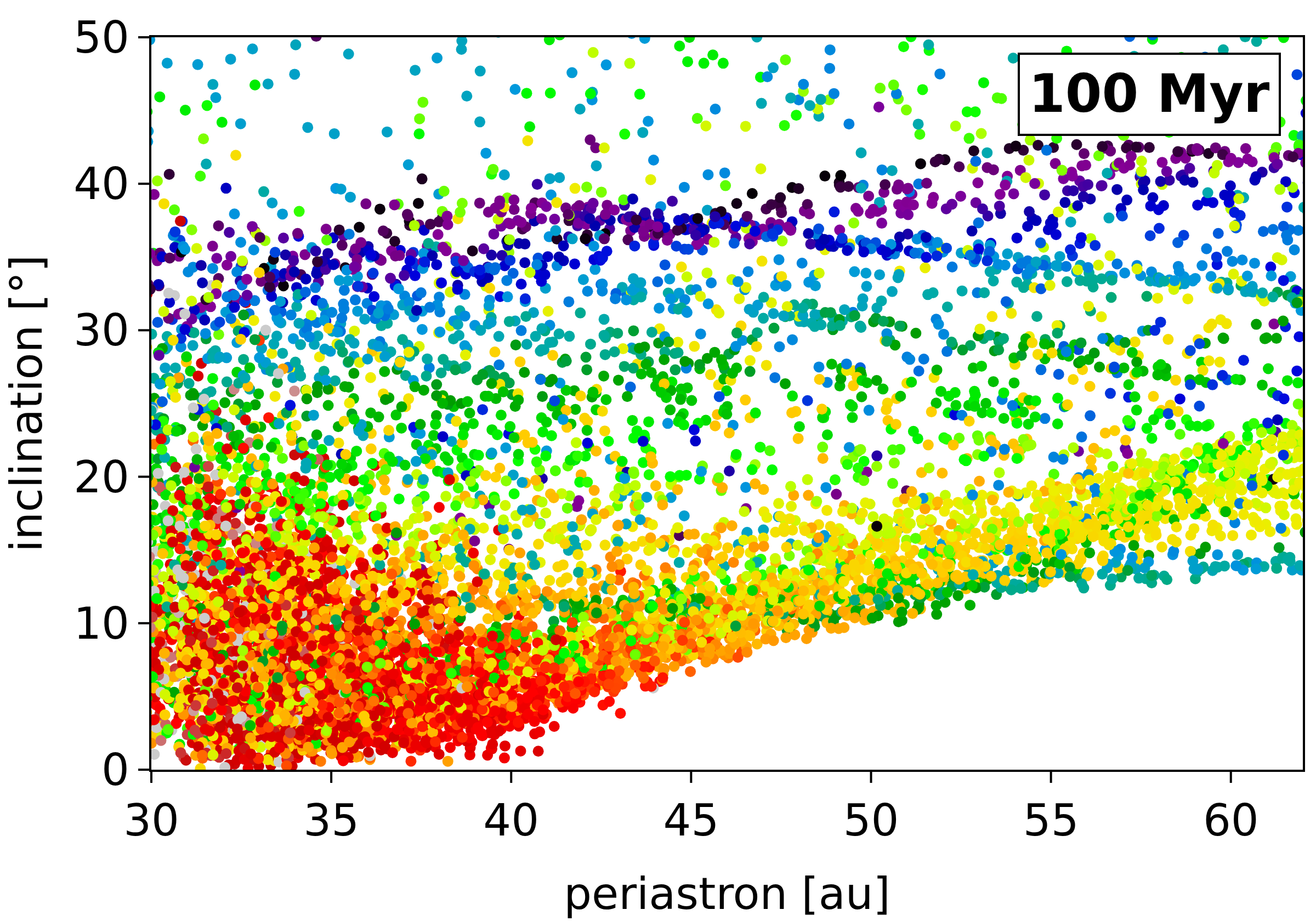}
    \textbf{(i)}
    \end{minipage}
    \\
    \begin{minipage}[b]{0.32\textwidth}
    \centering
    \includegraphics[width=\textwidth]{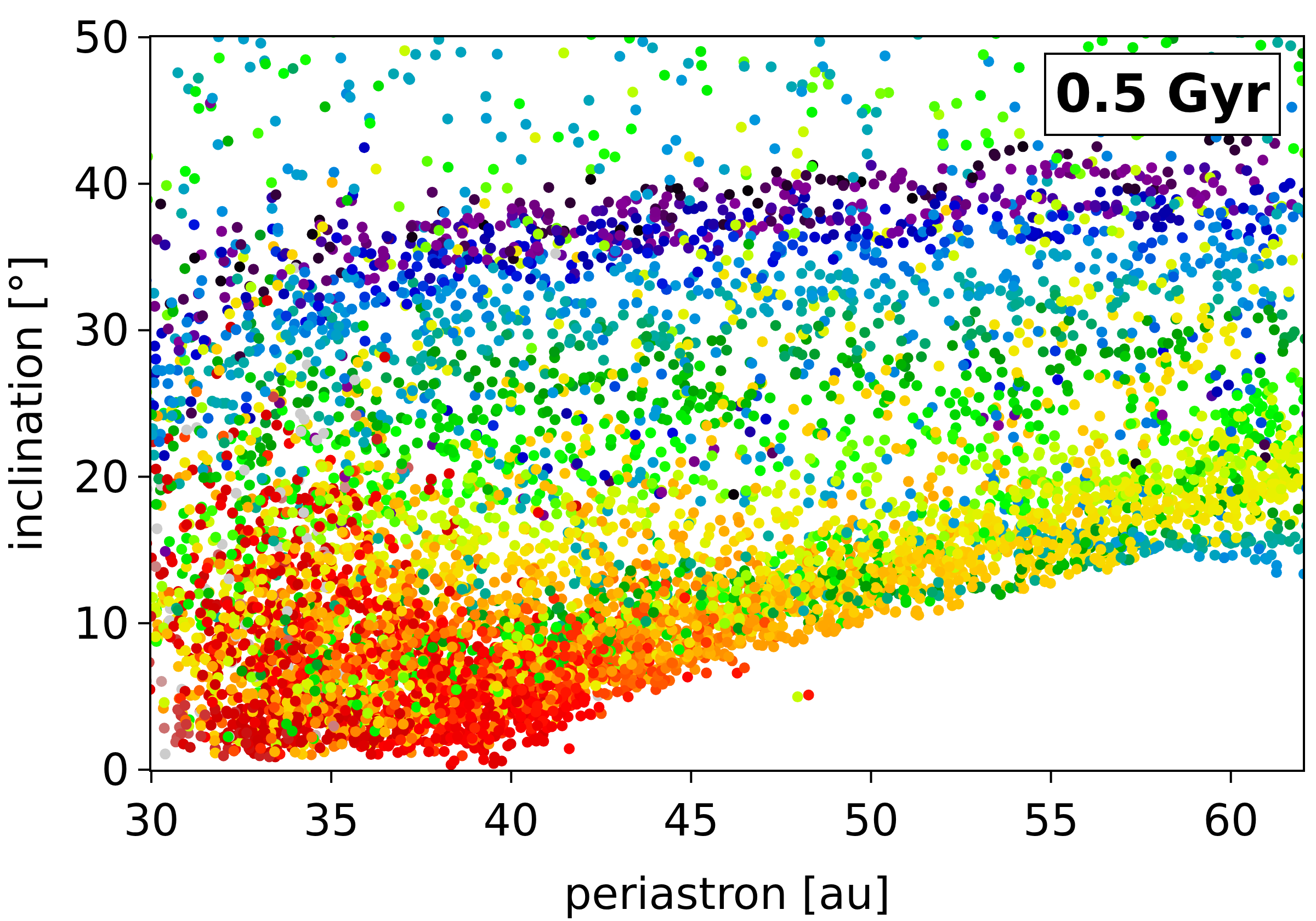}
    \textbf{(j)}
    \end{minipage}
    &
    \begin{minipage}[b]{0.32\textwidth}
    \centering
    \includegraphics[width=\textwidth]{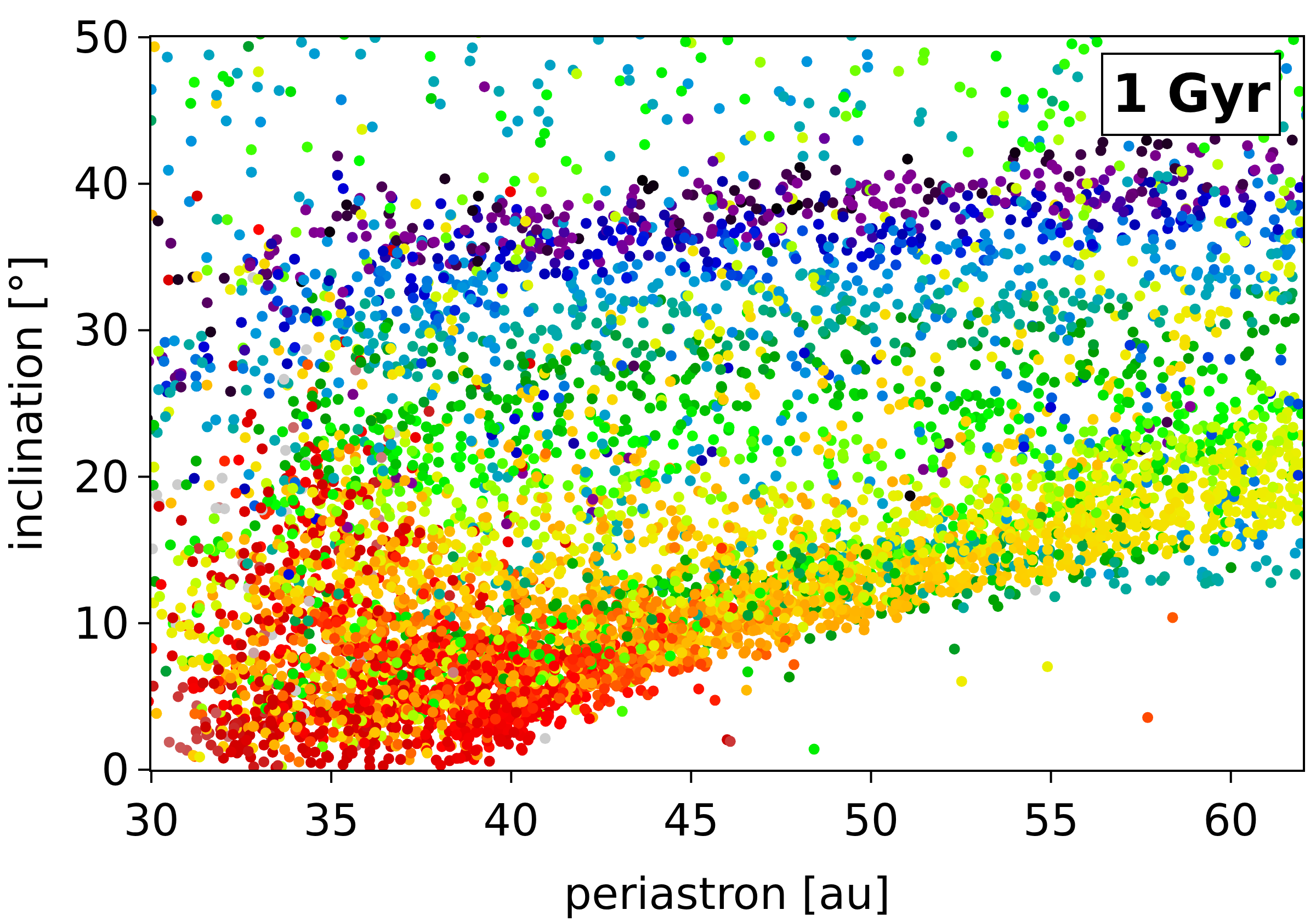}
    \textbf{(k)}
    \end{minipage}
    &
    \begin{minipage}[b]{0.32\textwidth}
    \centering
    \includegraphics[width=\textwidth]{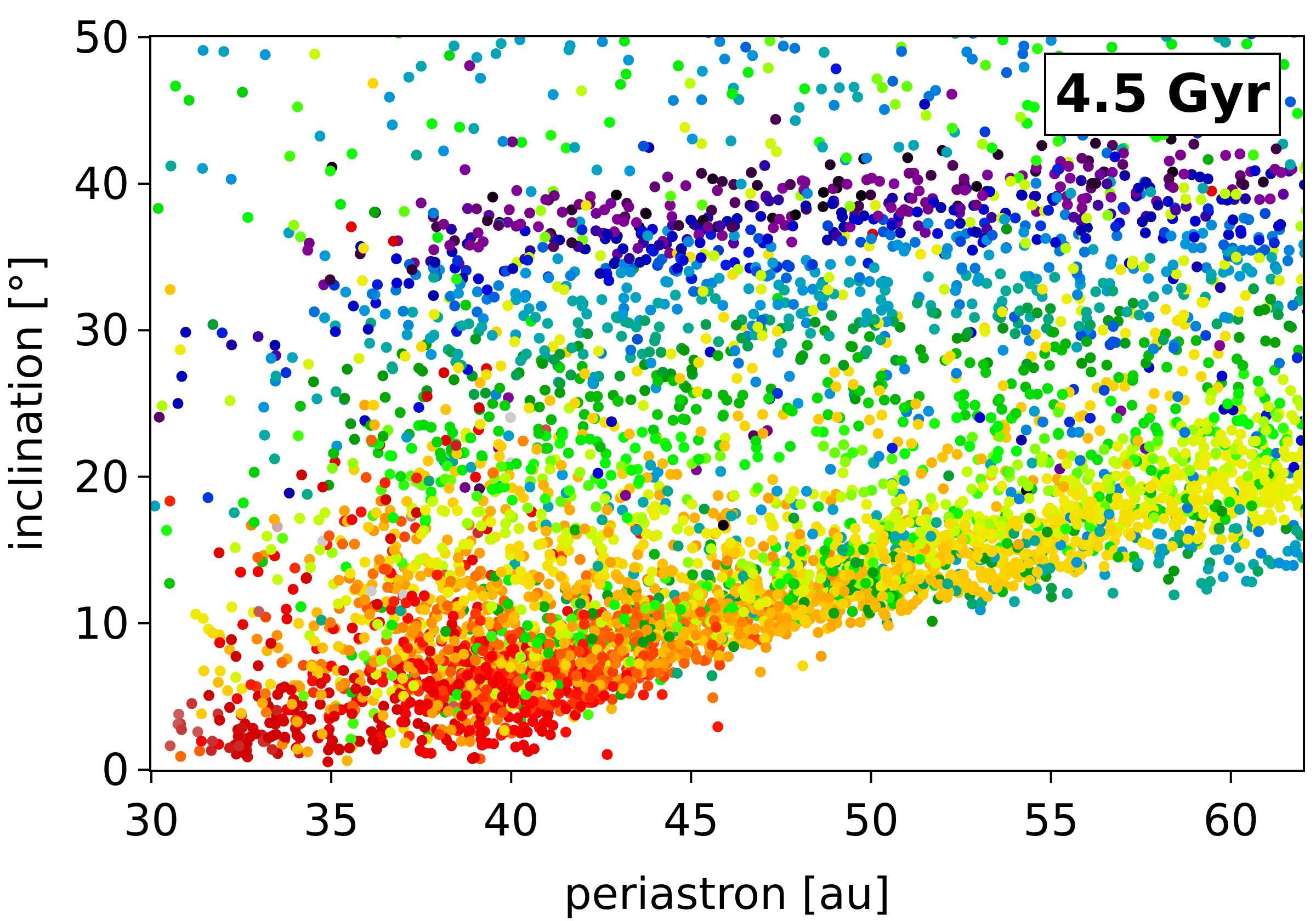}
    \textbf{(l)}
    \end{minipage}
    \\
\end{tabular}
\end{flushleft}
\caption{Long-term evolution of TNO dynamics. Panels (a)–(f): eccentricity space; panels (g)–(l): inclination space. Colours denote initial radial distance from the Sun before the flyby; see Fig.~\ref{fig:colours}(a).}
\label{fig:ecc_overview}
\end{figure*}

The pre-flyby disc was represented by 5$\times$10$^4$ massless tracer particles assuming an idealised thin disc. This resolution model allowed us to efficiently resolve the relevant dynamic groups. Debris discs around young stars typically range in size from 100~au to 500~au. We modelled disc sizes of 150~au and 300~au using an initially constant particle surface density to achieve high resolution in the outer parts of the disc. Post-processing of the data was carried out by assigning different masses to the particles, thereby modelling the actual mass density distribution. This approach reduced the computational costs of the study.

The long-term evolution study was performed using the GENGA code \citep{Grimm:2014, Grimm:2022}, employing its hybrid symplectic integrator \citep{Chambers:1999}. The GENGA code efficiently runs on GPU architectures; nevertheless, the computational cost remains relatively high. Therefore, we only accounted for the four giant planets in our simulations over the full 4.5~Gyr period. We use a fixed time step of 23~days to ensure an adequate resolution of trajectories as close as Mars' orbit. Recently, \citet{Rein:2025} found that Solar System simulations already converge with a time step of 32~days with this integrator. Therefore, our temporal resolution is sufficient for our type of study.

We also modelled the first 1~Gyr with the terrestrial planets included in the simulation. We found no significant differences between the results with and without terrestrial planets. Therefore, we neglected them in the simulation of the consecutive 3.5~Gyr.

\section{Results}
\label{sec:results}

\subsection{Ejected and relocated TNOs}

The dynamical evolution of the TNO population began with a significant depletion of bound TNOs during the initial 10~Myr to 100~Myr following the stellar flyby. This transient yet intense phase is characterised by the ejection of numerous TNOs (see Fig.~\ref{fig:particle_losses} top), with an ejection rate that exceeded current rates by a factor of 100 (Fig.~\ref{fig:particle_losses} bottom). After approximately 1~Gyr, the system reaches a relatively stable configuration, with a moderate ejection rate.

The loss of TNOs is most significant for bodies with periastron distances in the range 30~au~$< q <$~40~au. This is illustrated in Fig.~\ref{fig:particle_count}, which presents the temporal evolution of the test particle number density as a function of periastron distance after the flyby. TNOs lost from this region either have dramatically reshaped orbits or are ejected entirely, escaping the Solar System and becoming interstellar objects (ISOs).

\begin{table*}[t]
\begin{center}
\caption{TNO properties and abundance for a 150~au-sized solar disc. $N_x$ is the number of test particles immediately after the flyby, and $N$ refers to the total number of bound particles at a given time. Both values refer to test particles with 30.1~au~$< p <$~60~au, as this is the only region sufficiently covered by observations.}
\label{tab:TNO_properties}
      \begin{tabular}{llll|l|l|l|ll}
  \hline \hline
       \multicolumn{1}{l}{Name}    & \multicolumn{3}{l}{Class specifiers} & \multicolumn{5}{l}{Abundance} \\
\hline
      \multicolumn{4}{l}{}    & \multicolumn{1}{|l|}{t=0 Gyr}  & \multicolumn{1}{l|}{0.1 Gyr} & \multicolumn{1}{l|}{1 Gyr} &\multicolumn{2}{l}{t=4.5 Gyr}\\
              & $p$[au] & $e$ & $i$[$^o$] & $N_x/N$  & $N_x/N$ &  $N_x/N$ & $N_x/N$ & $N_x/N_{init}$\\ 
\hline 
Cold KB    & [30.1, 48] & $<$ 0.15 & $<$ 10    & 0.185 & 0.078  & 0.071 & 0.071  & 0.209 \\
Hot  KB    & [30.1, 48] &  $>$ 0.15 & [10, 35]  & 0.231 & 0.313  & 0.290  & 0.237 & 0.598\\
Detached   & [48, 60]   & $>$ 0.24 & $<$ 35    & 0.093 & 0.122  & 0.122  & 0.185 & 1.078 \\
Sedna-like & [60, 100]   & $>$ 0.6  &  $<$ 35   & 0.117 & 0.137  & 0.138  & 0.148 & 1.045\\
Inclined   & [30.1, 60] &  --      & [35, 90]  & 0.131 & 0.136  & 0.136  & 0.019 & 0.789\\
Retrograde & [30.1, 60] &  --      & $>$ 90    & 0.035 & 0.034  & 0.034  & 0.055 & 0.833\\
Inner      & [0.1, 29.5] &  --      & --       & 0.178 & 0.042  & 0.006  & 0.001 & 0.004\\
\hline    
\end{tabular}
\end{center}
\end{table*}

During the first 10~Myr, already 13\% of all test particles were ejected. By 4.5~Gyr, cumulative losses reached 63\% of the initial population in this periastron distance range. Consequently, the population of objects with periastron distances between 30~au and 40~au was originally at least three times larger than the present population in this range. Currently (August 2026), the Minor Planet Center (MPC) catalogue lists 3463 TNOs with 30.1~au$< q <$~40~au. Thus, this region would have contained at least 9360 TNOs just after the flyby, while nearly 6000 objects have been ejected since.

Figure~\ref{fig:ecc_overview} provides a more detailed illustration of how TNO loss depends on orbital parameters, showing snapshots of the population’s evolution in eccentricity Fig.~\ref{fig:ecc_overview}(a---f) and inclination Fig.~\ref{fig:ecc_overview}(g---l) as a function of periastron distance over 4.5~Gyr. Each object's colour reflects its pre-flyby origin in the disc (for details, see Sec.~\ref{sec:colours}).

The loss rate is influenced not only by the distance to Neptune but also by the TNOs' orbital eccentricities. Within the range 30~au~$< q <$~40~au, objects with low eccentricity ($e <$~0.2) are most severely depleted, with around 80\% of the initial population lost. In contrast, high-eccentricity TNOs ($e >$~0.4) in this region lose only 47\% of their members. Most survivors here are found in mean-motion resonance with Neptune, appearing as diagonal lines in the $e$--$q$ plot. Many non-resonant objects are ejected, but a significant fraction instead migrate onto more eccentric orbits -- depicted as high-eccentricity red points in Fig.~\ref{fig:ecc_overview}. This transition to higher eccentricities is especially pronounced 10~Myr -- 100~Myr after the flyby.

The objects' inclination also plays a significant role. High-inclination objects experience much less loss ($\simeq$~21\%) compared to low-inclination objects ($\simeq$~51\%). According to the MPC catalogue, there are 1092 objects with $i <$~10° and 30~au~$< q <$~40~au. Immediately after the flyby, there should have been more than 5500 objects in this parameter space. A considerable fraction that initially had 30~au~$< q <$~40~au is scattered to higher inclinations $i$, although very few are pushed beyond $i >$~20°.

In contrast, TNOs characterised by perihelion distances $q >$~40~au, eccentricities $e >$~0.5, and inclinations $i >$~10° are rarely ejected and exhibit negligible variation in these orbital parameters, even over a 4.5~Gyr evolutionary timescale. Notably, Sedna-like objects -- including Sedna, Biden, Leleākūhonua, and Ammonite -- occupy this region of parameter space and remain dynamically stable. Consequently, Sedna-like objects serve as robust tracers for constraining early Solar System dynamical models, as their orbital elements have undergone minimal alteration throughout Solar System history. Their properties are also critical for determining the best-fit flyby scenarios.

Neptune is expected to create a gap of $\pm$ 3 Hill radii {\citep{Pirani:2021}, with $R_H \approx a (1 - e) \sqrt[3]{m/ 3 M} \approx 0.78$~au. Figure ~\ref{fig:ecc_overview}} (b) shows that, for low-eccentric objects, most of this clearance is already achieved after 10~Myr for low-inclination objects. However, for high-inclination objects, the clearing takes much longer.

\subsection{TNO dynamical groups}

TNOs are commonly classified based on their orbital elements, such as semi-major axis, eccentricity, and inclination. Various classification schemes exist, but most differentiate between cold and hot classical Kuiper Belt objects (KBOs) and scattered and detached TNOs \citep{Gladman:2021}. In addition, subpopulations such as Sedna-like objects, high-inclination TNOs, and retrograde TNOs warrant particular attention due to their unique dynamical histories and formation pathways. Although only a limited number of these atypical objects have been identified to date, comprehensive models of TNO formation and evolution must account for their observed properties and population statistics. The definitions adopted in this work are summarised in Tab.~\ref{tab:TNO_properties}.

\begin{figure}
\centering
    \includegraphics[width=0.48\textwidth]{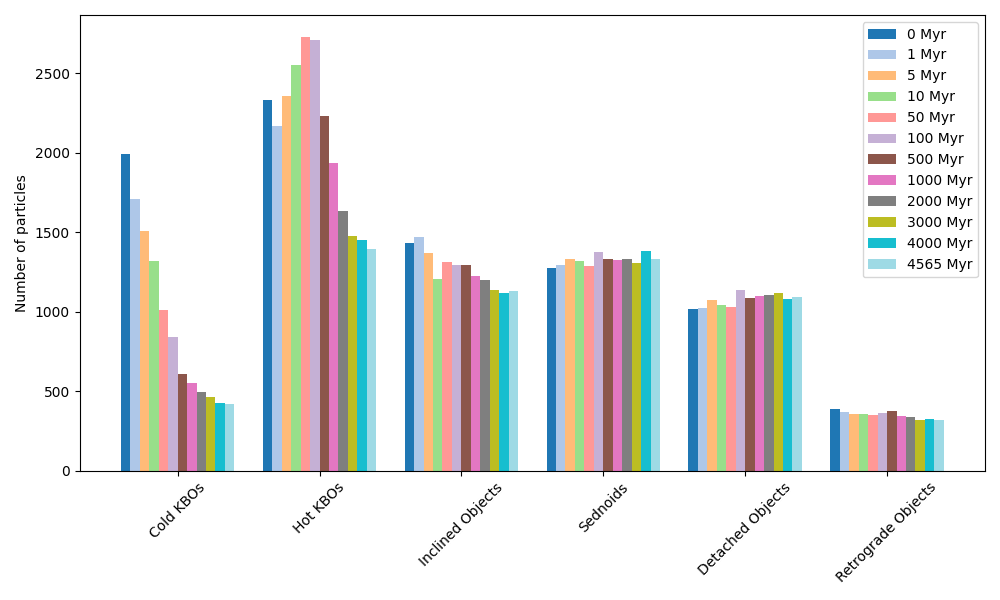}
     \includegraphics[width=0.48\textwidth]{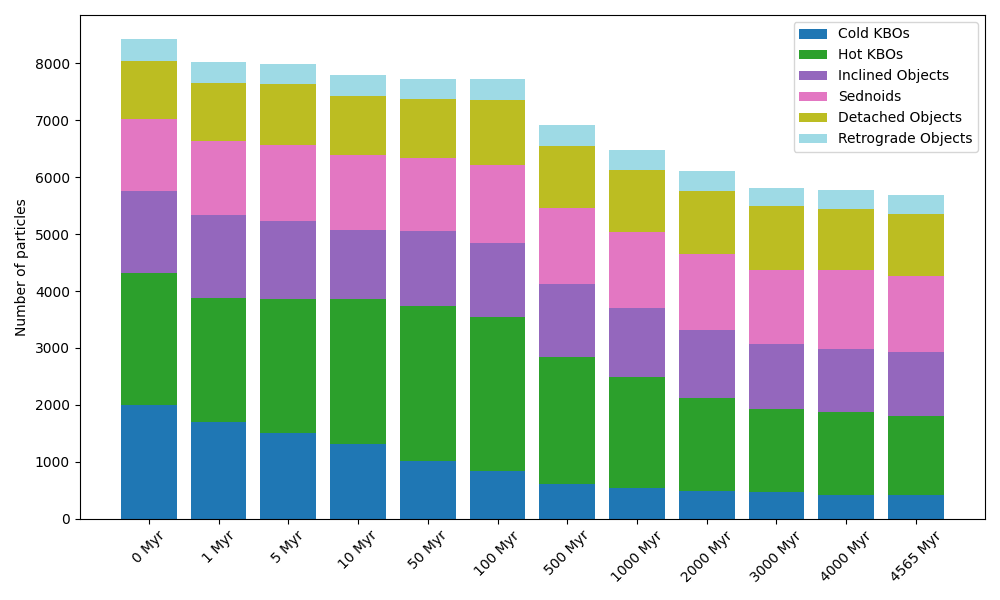}
\caption{Evolution of the number of TNOs in different dynamical families as a function of time elapsed after the flyby. Top: membership in the different dynamical groups as a function at different times. Bottom: relative contribution of different dynamical groups at various times after the flyby.}
\label{fig:losses-by-family}
\end{figure}

Figure~\ref{fig:losses-by-family} top presents the temporal evolution of the TNO dynamical families over multiple epochs. Here we look at all test particles with $q <$~100~au. The results demonstrate that three populations -- the cold classical KBOs, the hot classical KBOs, and the scattered disc objects -- undergo substantial depletion, with the cold classical population exhibiting the highest fractional loss. In contrast, the detached, Sedna-like, and retrograde TNO subpopulations exhibit remarkable dynamical stability. Specifically, the retrograde TNO population exhibits only a marginal decrease, Sedna-like objects maintain nearly constant numbers, and the detached population even shows a slight increase over time.

These distinct evolutionary trends also alter the relative population fractions among the various dynamical classes of TNOs (see Fig.~\ref{fig:losses-by-family} bottom). Initially, the cold and hot classical populations together constitute $\approx$~50\% of the total TNO population; however, this fraction declines to roughly one-third over time. The cold classical KBOs, in particular, transition from being the second-most populous subgroup to a relatively minor component within the overall TNO population.

\begin{figure*}
\centering
    \begin{minipage}[b]{0.31\textwidth}
    \centering
    \includegraphics[width=\textwidth]{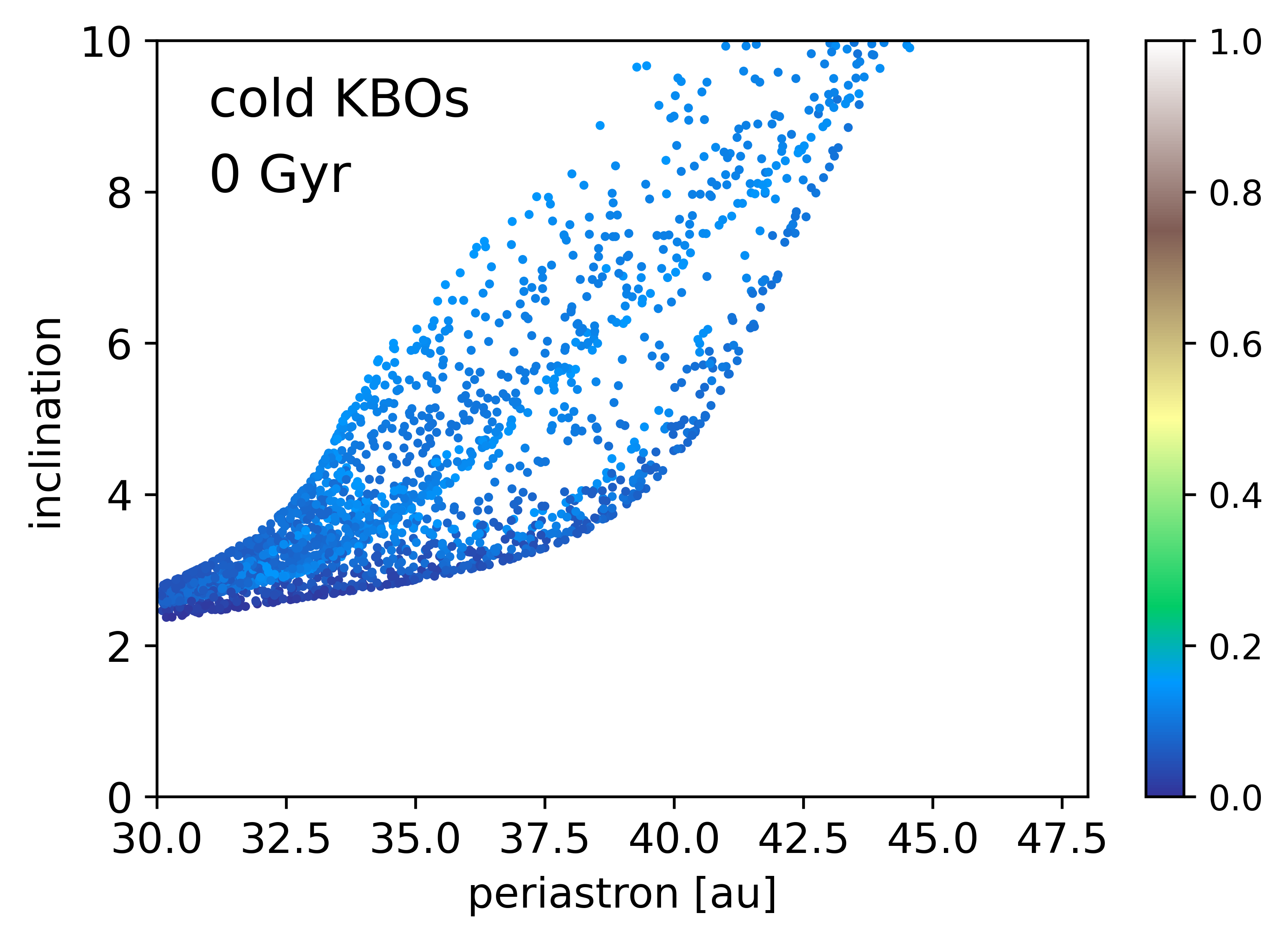}
  \end{minipage}
\centering
    \begin{minipage}[b]{0.31\textwidth}
    \centering
    \includegraphics[width=\textwidth]{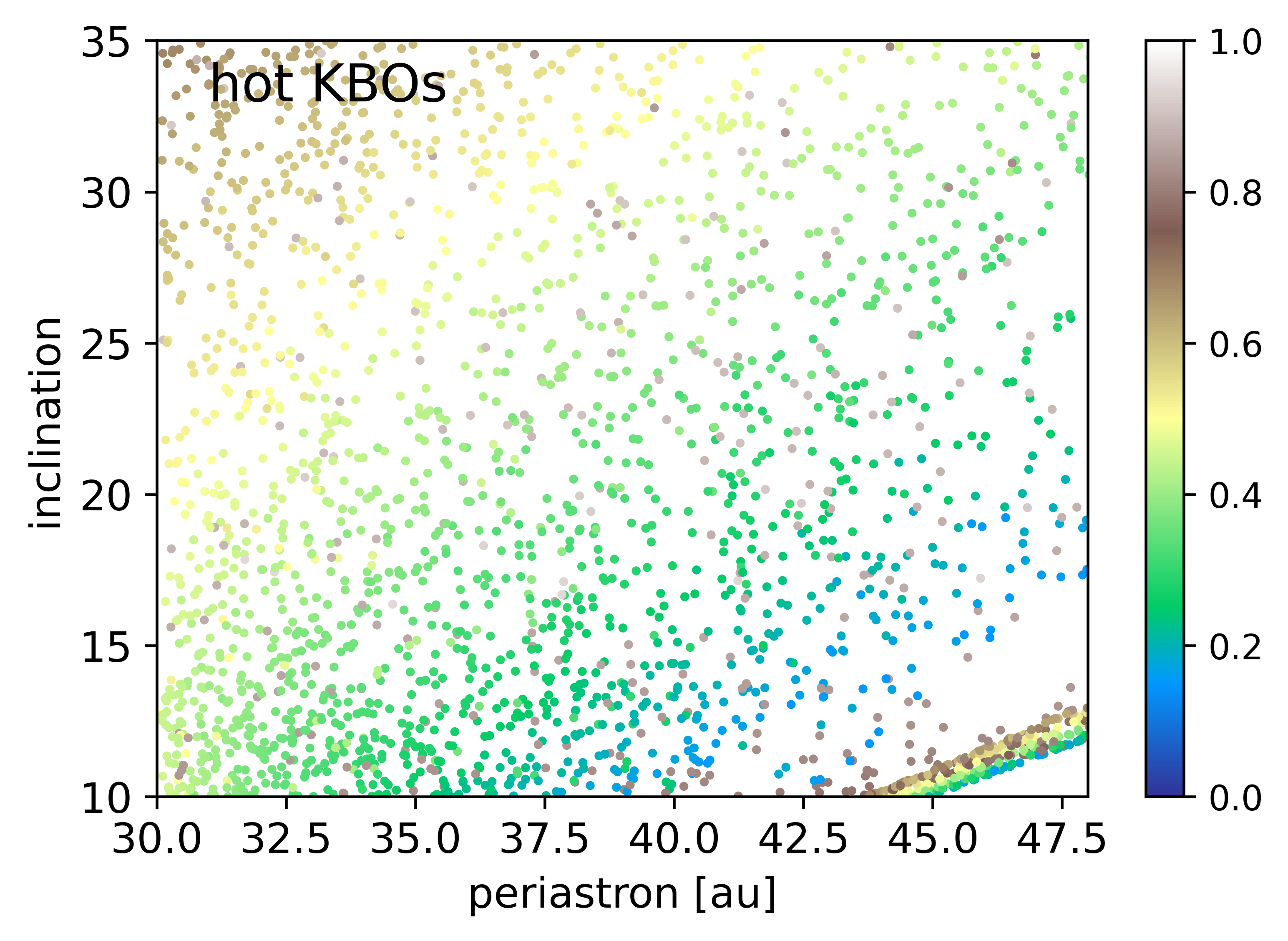}
  \end{minipage}
\centering
   \begin{minipage}[b]{0.31\textwidth}
   \centering
    \includegraphics[width=\textwidth]{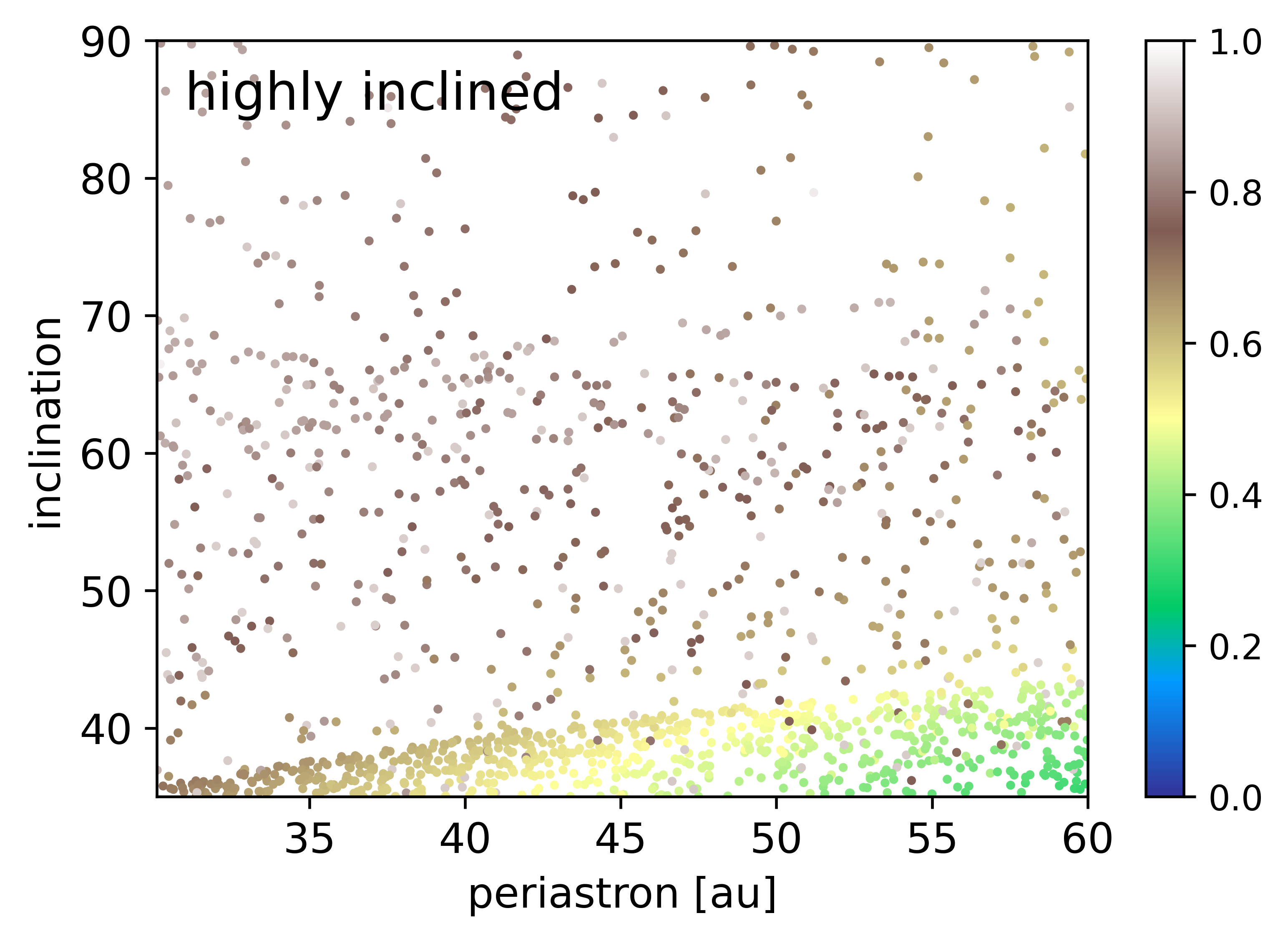}
  \end{minipage}

\centering
    \begin{minipage}[b]{0.32\textwidth}
    \centering
    \includegraphics[width=\textwidth]{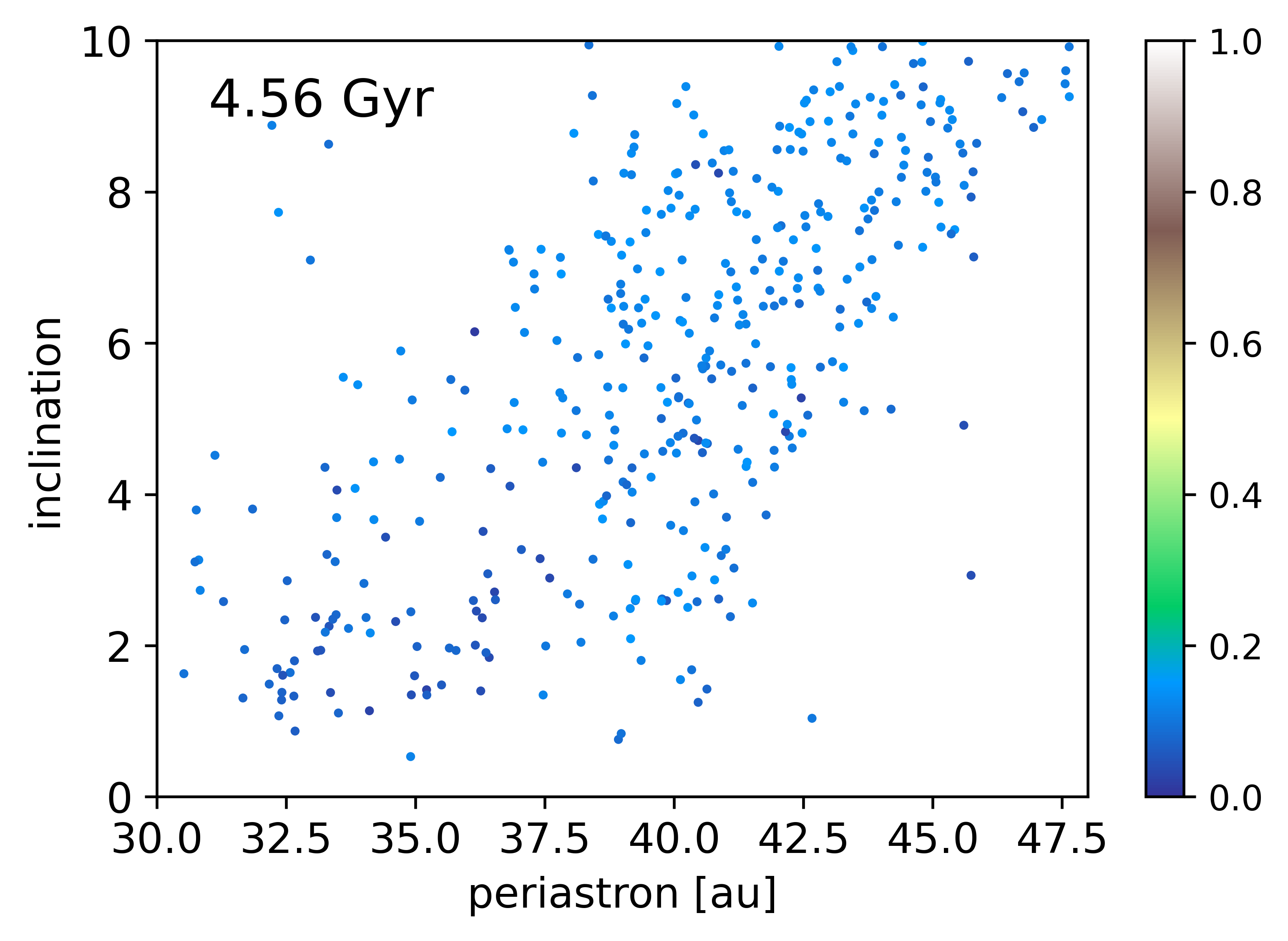}
    \textbf{(a)}
  \end{minipage}
\centering
    \begin{minipage}[b]{0.32\textwidth}
    \centering
    \includegraphics[width=\textwidth]{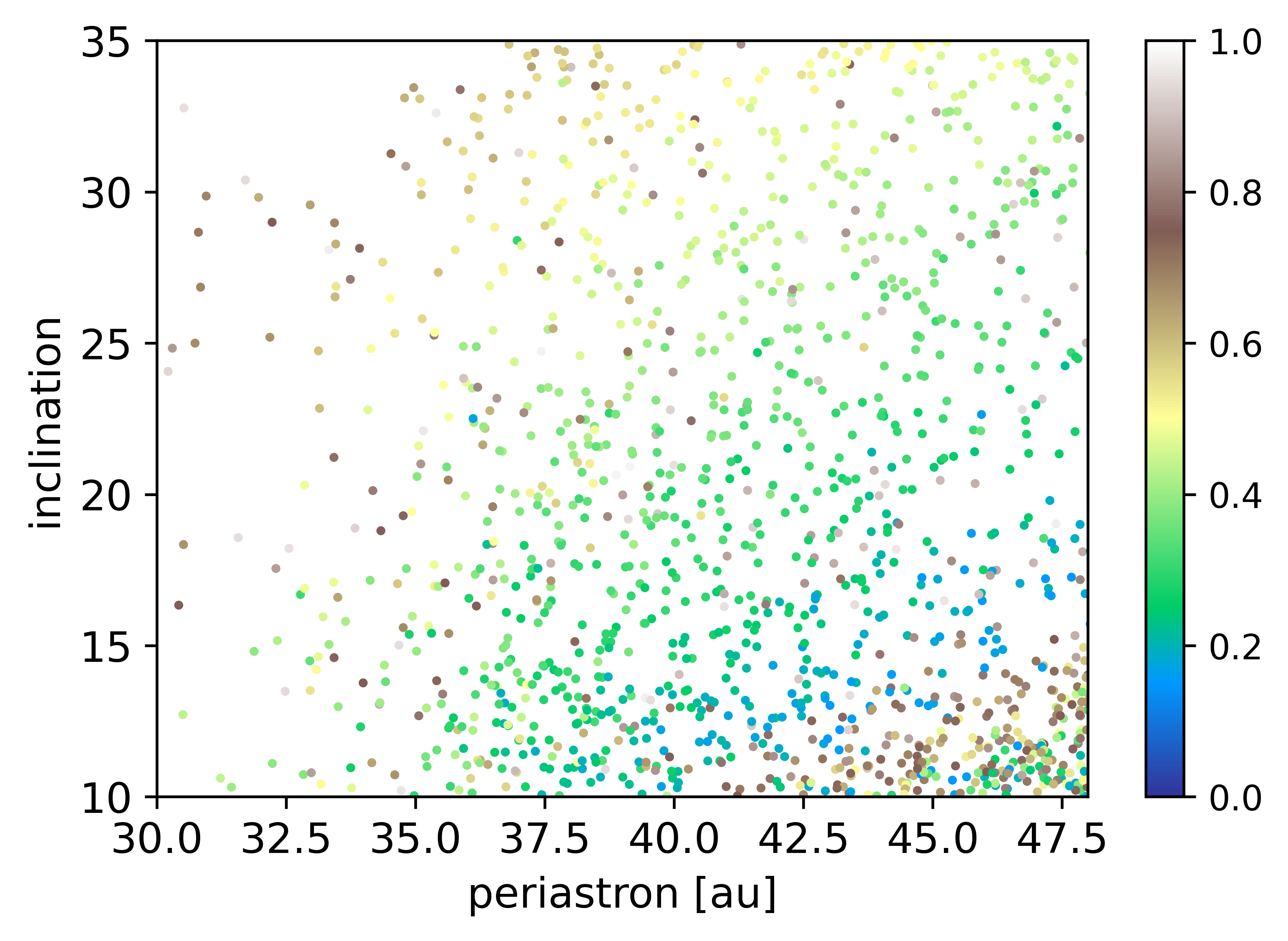}
    \textbf{(b)}
  \end{minipage}
\centering
    \begin{minipage}[b]{0.32\textwidth}
    \centering
    \includegraphics[width=\textwidth]{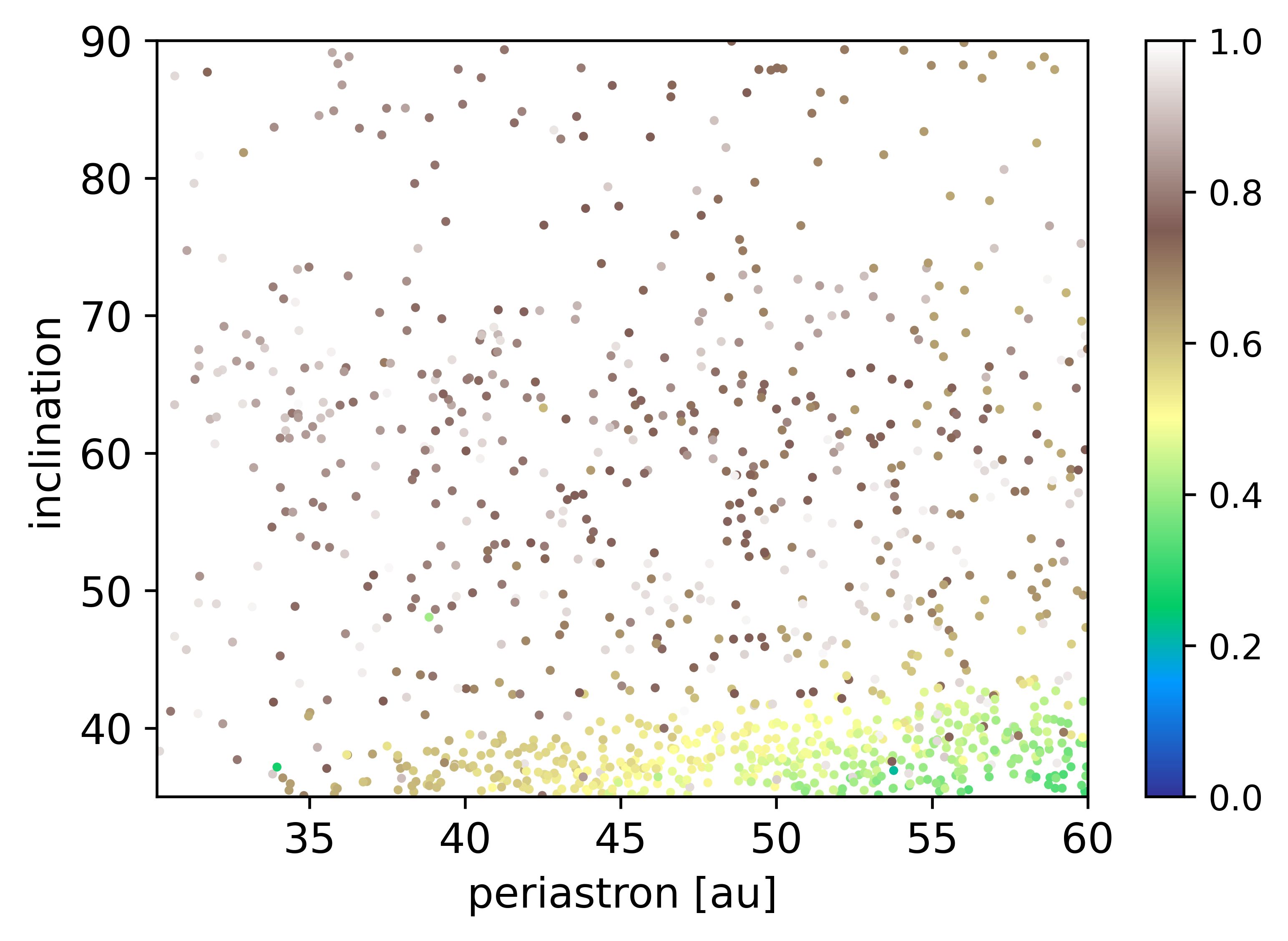}
    \textbf{(c)}
  \end{minipage}
  
\caption{Evolution of cold and hot KBOs and high-inclination TNOs. Shown are the inclination vs the periastron distance of the test particles in our simulations. The colour indicates their eccentricity. The top row shows the dynamic properties directly after the flyby, the bottom row the same after 4.56~Gyr of dynamic evolution for a) cold and b) hot KBOs and c) high-inclination TNOs. For the definition of these dynamic groups, see Tab.~\ref{tab:TNO_properties}.}
\label{fig:families_chd}
\end{figure*}

Next, we systematically quantify the dynamical evolution of each TNO subpopulation. Figures~\ref{fig:families_chd} and \ref{fig:Families_sdr} present the distribution of orbital inclination as a function of perihelion distance for the respective dynamical classes. Colour coding in these figures represents particle eccentricity. The upper panels correspond to the orbital configuration immediately following the stellar flyby, while the lower panels depict the state after 4.5~Gyr of dynamical evolution.

\subsubsection{Cold Kuiper belt population}

Among all dynamically distinct families, the cold KBOs are most strongly affected by gravitational interactions with Neptune. Figure~\ref{fig:families_chd}a presents a comparison between the cold KBO population immediately after the flyby (top panel) and its distribution 4.5~Gyr later. Initially, cold KBOs occupy a well-defined region of parameter space, characterised by perihelion distances $q <$~33~au and the absence of objects with inclinations $i <$~2°. During 4.5~Gyr of dynamical evolution, this configuration undergoes a significant evolution.

By 4.5~Gyr, the total number of cold KBOs has decreased significantly, and most areas of the parameter space contain cold KBOs. In particular, there are a considerable number of cold KBOs with $i <$~2°. The density of cold KBOs is now highest for $q >$~38~au. In some way, the observed ``inner edge'' of the cold KBO population is a result of the ejection of particles inside $\approx$~38~au. In summary, after 4.5~Gyr, the simulated cold KBO population matches the observed one much more closely.

Between 30~au and 50~au, $\approx$~21\% of the cold TNOs become unbound. Some particles remain bound, but obtain higher eccentricities or higher inclinations. Those excited, formerly cold, TNOs are visible in Fig. \ref{fig:ecc_overview} as red particles with eccentricities $e <$ 0.5 and inclination $i >$~10°.

\subsubsection{Hot Kuiper belt population}

The initially substantial population of dynamically hot KBOs underwent a significant depletion over the evolutionary 4.5~Gyr. In particular, objects with perihelion distances $q <$~35~au were predominantly eliminated, with only a small fraction persisting, typically on highly eccentric orbits (0.4~$< e <$~1.0) (see Fig.~\ref{fig:families_chd}b). Likewise, most objects with inclinations between 25° and 35° were lost, and the few that remain mostly have very elongated orbits (0.4~$< e <$~1.0). After 4.5~Gyr, there are nearly no TNOs with 30.1~au~$< q <$~35~au and 30°~$< i <$~35° left.

However, certain regions also exhibit an increase in population density over time. Especially, the previously pronounced gap located in the lower right quadrant of the diagram (see Fig.~\ref{fig:families_chd}b, top panel) becomes entirely populated with TNOs after 4.5~Gyr (bottom panel). The area that was initially devoid of objects -- defined by 45~au~$< q <$~48~au and 12°~$< i <$~17° -- develops a significant population of TNOs over 4.5~Gyr, many of which are characterised by highly eccentric orbits ($e >$~0.6).

The presence of distinct diagonal colour gradients indicates a linear correlation between the perihelion distance, the inclination, and the eccentricity within the hot KBOs. This relationship is particularly preserved for TNOs with $e <$~0.4. This pattern provides a testable prediction for future observational studies.

\subsubsection{High-inclination objects}

Figure~\ref{fig:families_chd}c illustrates the temporal evolution of high-inclination TNOs (35°~$< i <$~90°). Again, TNOs with perihelion distances $q <$~35~au exhibit the greatest dynamical modification. Besides, the subpopulation characterised by both high eccentricities and moderate inclinations (35°~$< i <$~40°), which appears as a brown stripe in the upper panel of Fig.~\ref{fig:families_chd}c, is entirely depleted after 4.5~Gyr of dynamical evolution. In contrast, TNOs with inclinations $i <$~45° predominantly persist on highly eccentric orbits throughout the simulation.

\subsubsection{Detached, Sedna-like, and retrograde TNOs}

The detached and Sedna-like objects show negligible dynamical evolution throughout the integration timespan (see Fig.~\ref{fig:Families_sdr}a and b). This invariance is consistent with their large semi-major axes ($a >$~48~au), which precludes scattering interactions with the known giant planets. Furthermore, their orbits place them well beyond the present-day gravitational influence of Neptune, resulting in long-term orbital stability over Gyr. Our numerical simulations thus confirm theoretical expectations regarding the dynamical stability of these two populations. This outcome is particularly reassuring, as these objects serve as key constraints when evaluating plausible stellar flyby scenarios.

Nevertheless, discernible modifications are observed in regions characterised by a high concentration of TNOs. The initially distinct dynamical features (top) become increasingly diffuse (bottom). At present, it remains uncertain whether this phenomenon reflects genuine physical processes or is due to numerical artefacts inherent to the simulation methodology. More targeted investigation is required to resolve this ambiguity.

Similarly, the retrograde population is nearly invariant in the number of members (see Fig.~\ref{fig:losses-by-family} top), and their orbital parameters remain largely stable throughout the temporal evolution (see Fig.~\ref{fig:Families_sdr}c). Even within the periastron interval 30~au~$< q <$~35~au, only a few objects are lost. The reason for such stability is that the majority of objects within this periastron range possess orbital inclinations in the range 90°~$< i <$~110°, where Neptune's influence is negligible.

In conclusion, the dynamical stability observed in the orbits of detached, Sedna-like, and retrograde TNOs indicates that these populations are optimal for constraining stellar flyby parameters in numerical simulations that do not encompass the full post-flyby dynamical evolution. For all other TNO populations, rigorous quantitative comparisons require careful consideration of their subsequent dynamical evolution, at least in an approximate manner.

\subsubsection{Injected TNOs - Centaurs}

\begin{figure}
\centering
    \begin{minipage}[b]{0.48\textwidth}
    \centering
    \includegraphics[width=\textwidth]{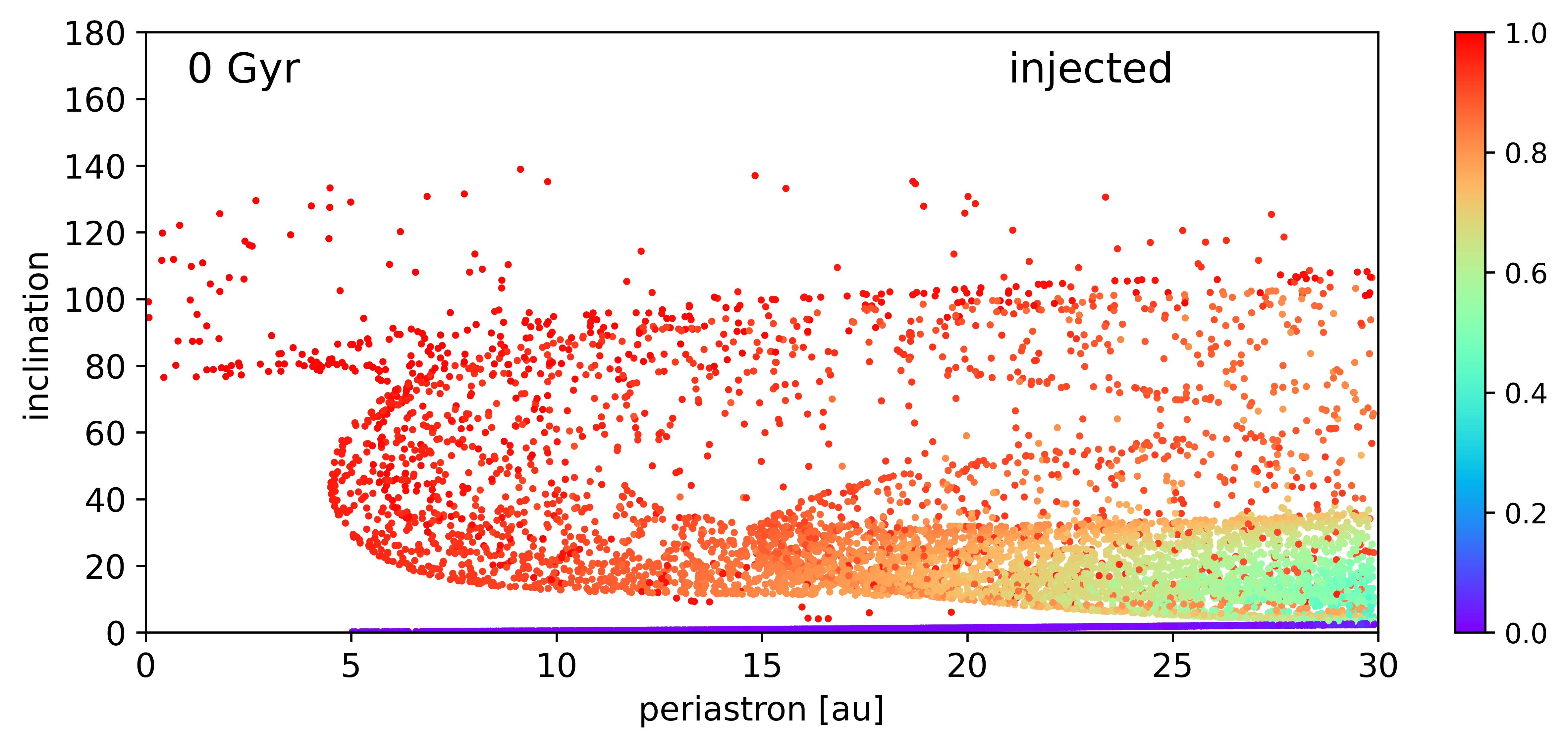}
  \end{minipage}
\centering
    \begin{minipage}[b]{0.48\textwidth}
    \centering
    \includegraphics[width=\textwidth]{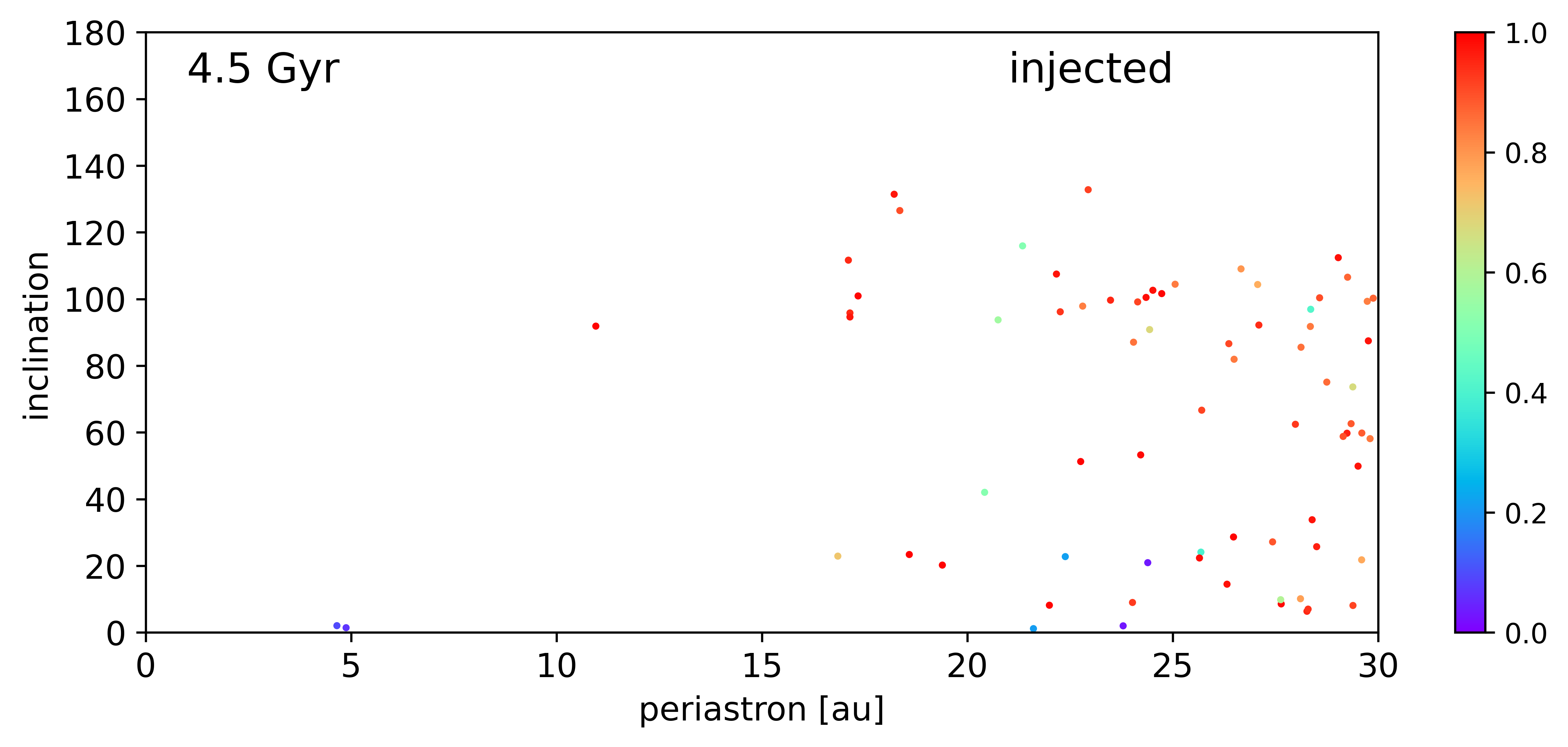}
  \end{minipage}

\caption{Evolution of TNOs injected into the planet area. Shown are the inclination vs the periastron distance of the test particles in our simulations. The colour indicates their eccentricity. The top row shows the dynamic properties directly after the flyby, the bottom row the same after 4.56~Gyr of dynamic evolution.}
\label{fig:Families_injected}
\end{figure}

A substantial fraction of TNO are injected into the region occupied by the giant planets. Of these, 99.1\% are subsequently ejected over the ensuing 4.5~Gyr, whereas only 0.9\% persist on these timescales (see Fig.~\ref{fig:Families_injected}). The surviving injected TNOs contribute to the present-day Centaur population and predominantly occupy highly eccentric orbits, many crossing Neptune's orbit. The results further suggest that TNOs injected onto retrograde orbits exhibit an enhanced probability of long-term survival compared to their prograde Centaur counterparts.

Nonetheless, in the simulations a significant fraction of objects with low inclinations ($i <$~40°) and long-periods ($a >$~100~au with perihelia of $q <$~30~au) is present after 4.5~Gyr. Such objects are observed, and \citet{Batygin:2024} claimed that their existence is further evidence for the existence of a ninth planet. They base this claim on simulations without a ninth planet underrepresenting this population. Our findings indicate that the existence of a ninth planet is not the sole mechanism capable of producing Centaurs with these orbital characteristics. The proposed stellar flyby scenario likewise generates such objects. Consequently, the presence of these objects serves as evidence that any external perturbation -- be it from a distant planet or a passing star -- gives rise to Centaurs with these orbits, persisting even after 4.5~Gyr of dynamical evolution.

Although there is qualitative agreement with the observed Centaurs, quantitative analysis is not yet possible. Such a comparison would necessitate higher-resolution numerical simulations.

\section{Comparison with observed TNOs}

Next, we investigate how our model compares after 4.5~Gyr of evolution with the observed TNO population. Any such comparison is complicated by the fact that the observed TNO population is affected by numerous observational biases. First, distant and small TNOs are to a large degree missing in current catalogues. Similarly, TNOs on highly eccentric orbits are underrepresented, as are high-inclination objects. In total, the TNOs discovered so far likely represent only a fraction ($<<$~10\%) of the total population  [for a more detailed discussion, see \citet{Bannister:2018,Gladman:2021,Bernardinelli:2022}]. Therefore, a direct comparison between observations and simulations is subject to significant limitations. However, while we cannot account for the undetected TNOs, any simulation must reproduce the properties of the known TNOs, such as the KBOs, Sedna-like, and retrograde populations.

\begin{figure}
\centering
  \begin{minipage}[b]{\linewidth}
    \centering
    \includegraphics[width=\textwidth]{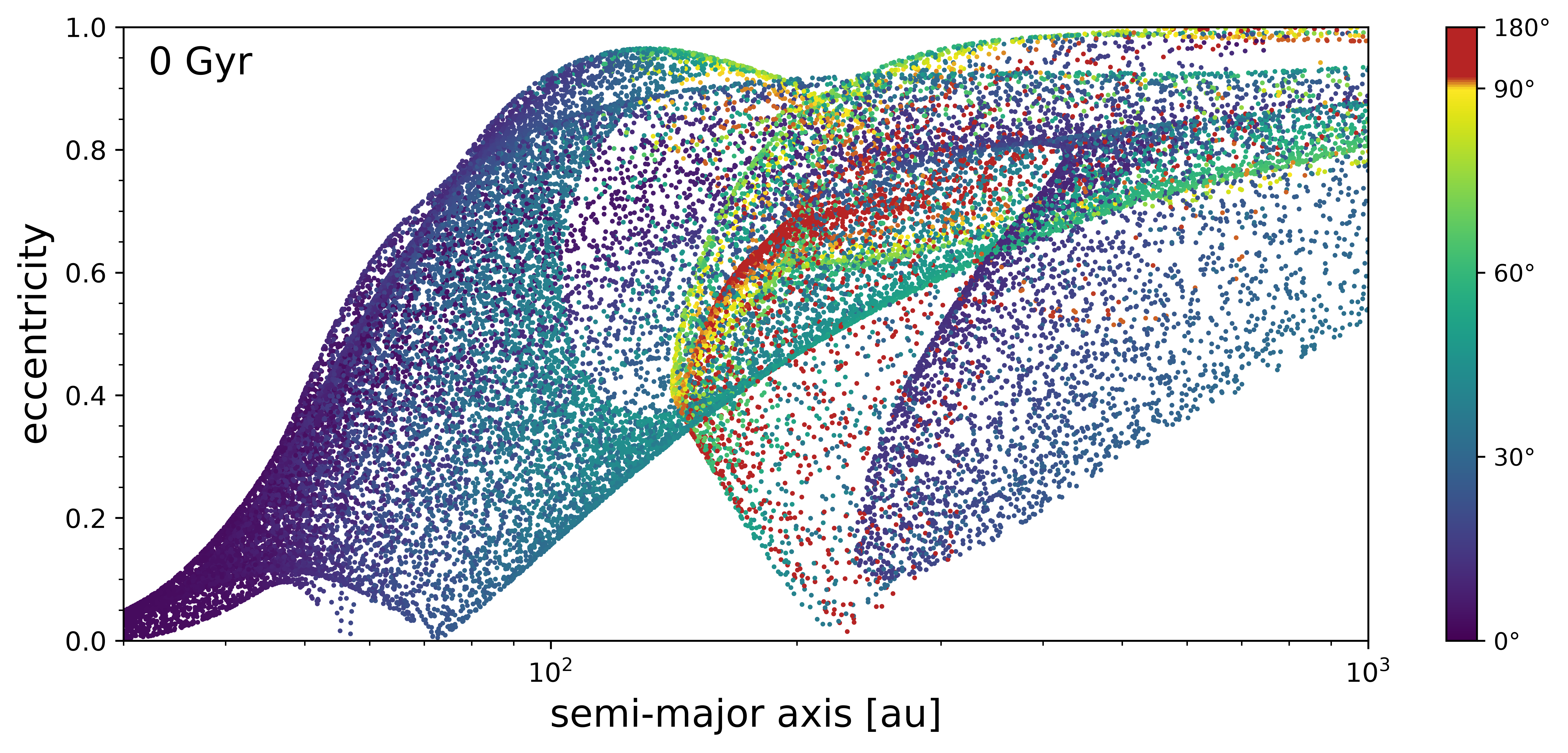}
    \textbf{(a)}
  \end{minipage}

  \centering
  \begin{minipage}[b]{\linewidth}
    \centering
    \includegraphics[width=\textwidth]{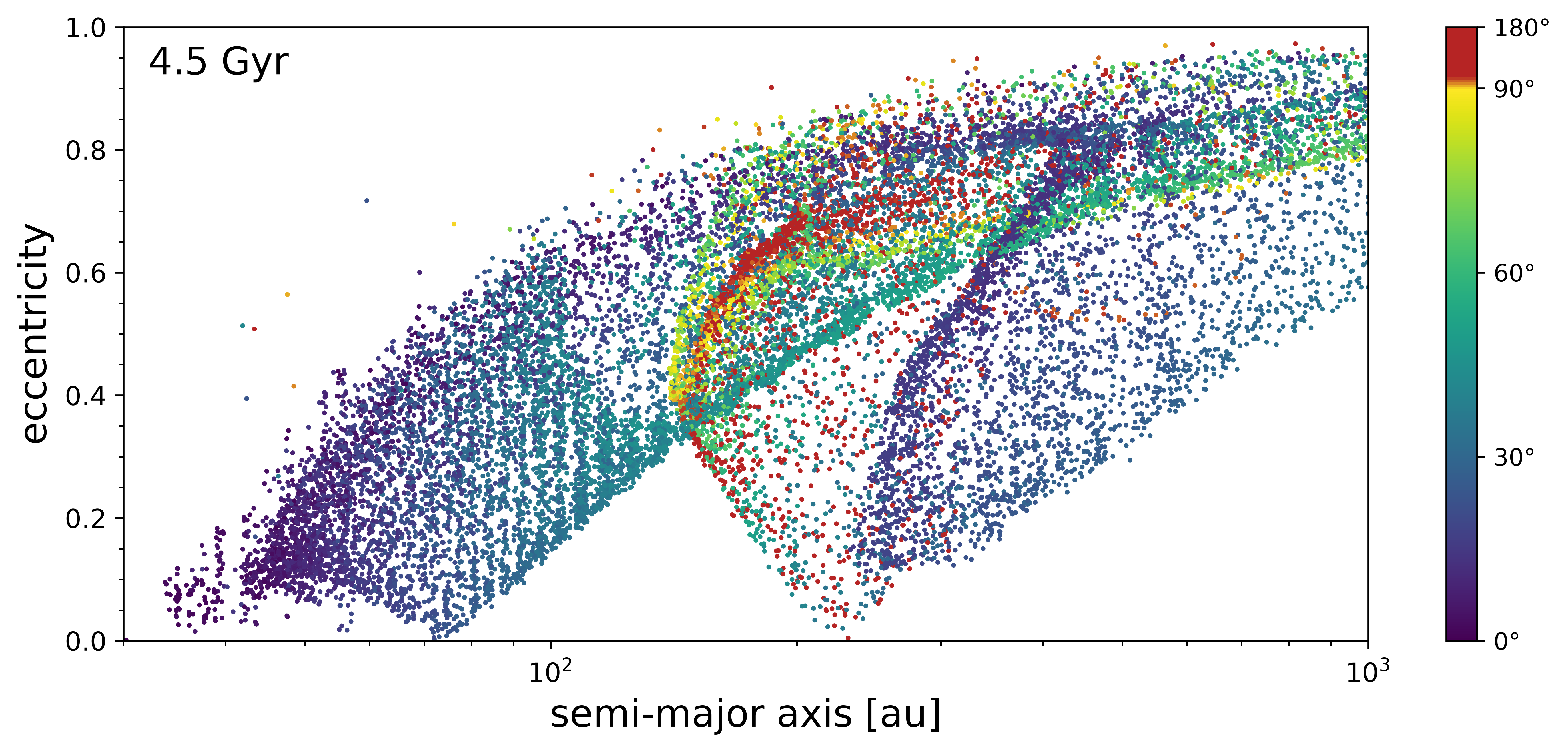}
    \textbf{(b)}
  \end{minipage}

 \centering
  \begin{minipage}[b]{\linewidth}
    \centering
    \includegraphics[width=\textwidth]{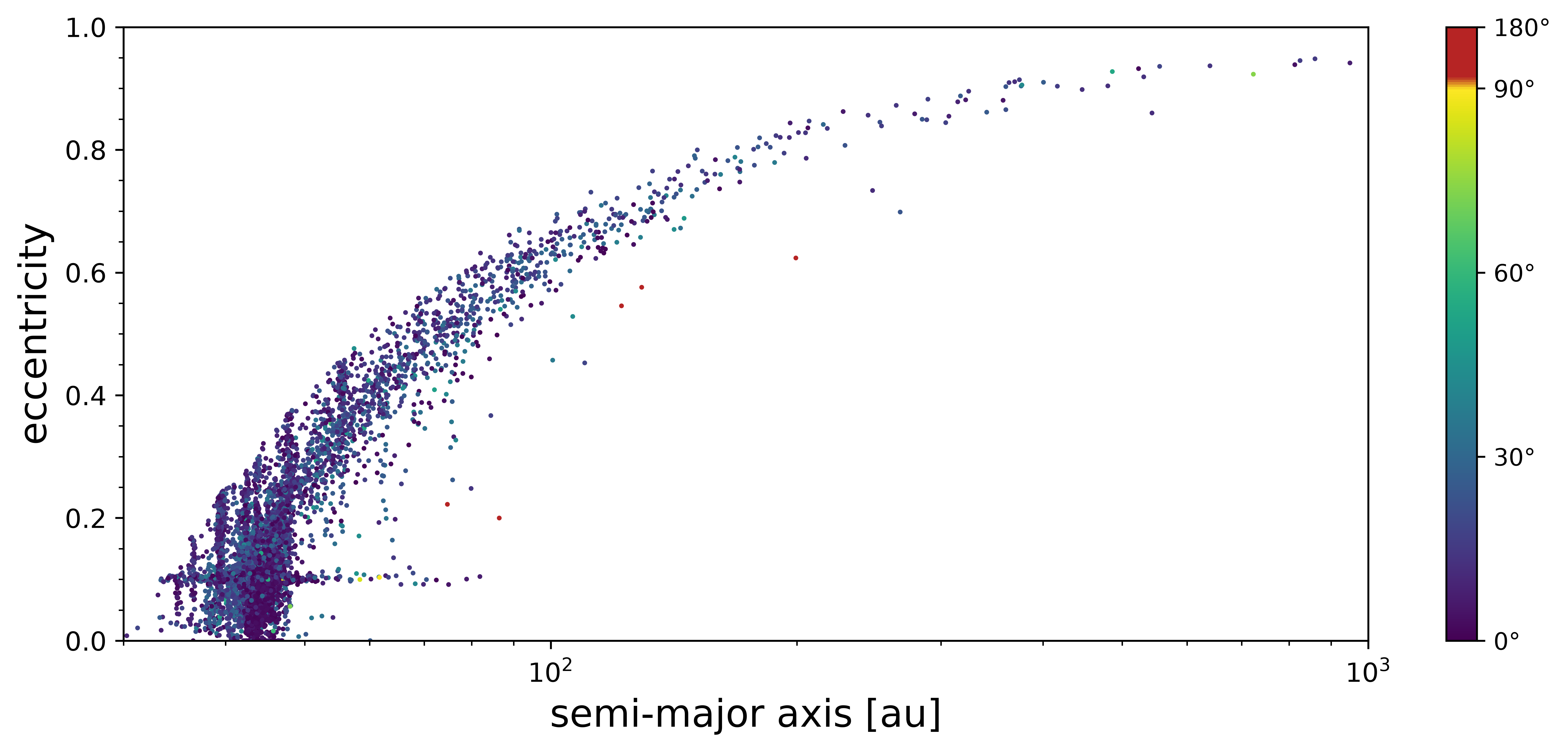}
    \textbf{(c)}
  \end{minipage}

 \centering
  \begin{minipage}[b]{\linewidth}
    \centering
    \includegraphics[width=\textwidth]{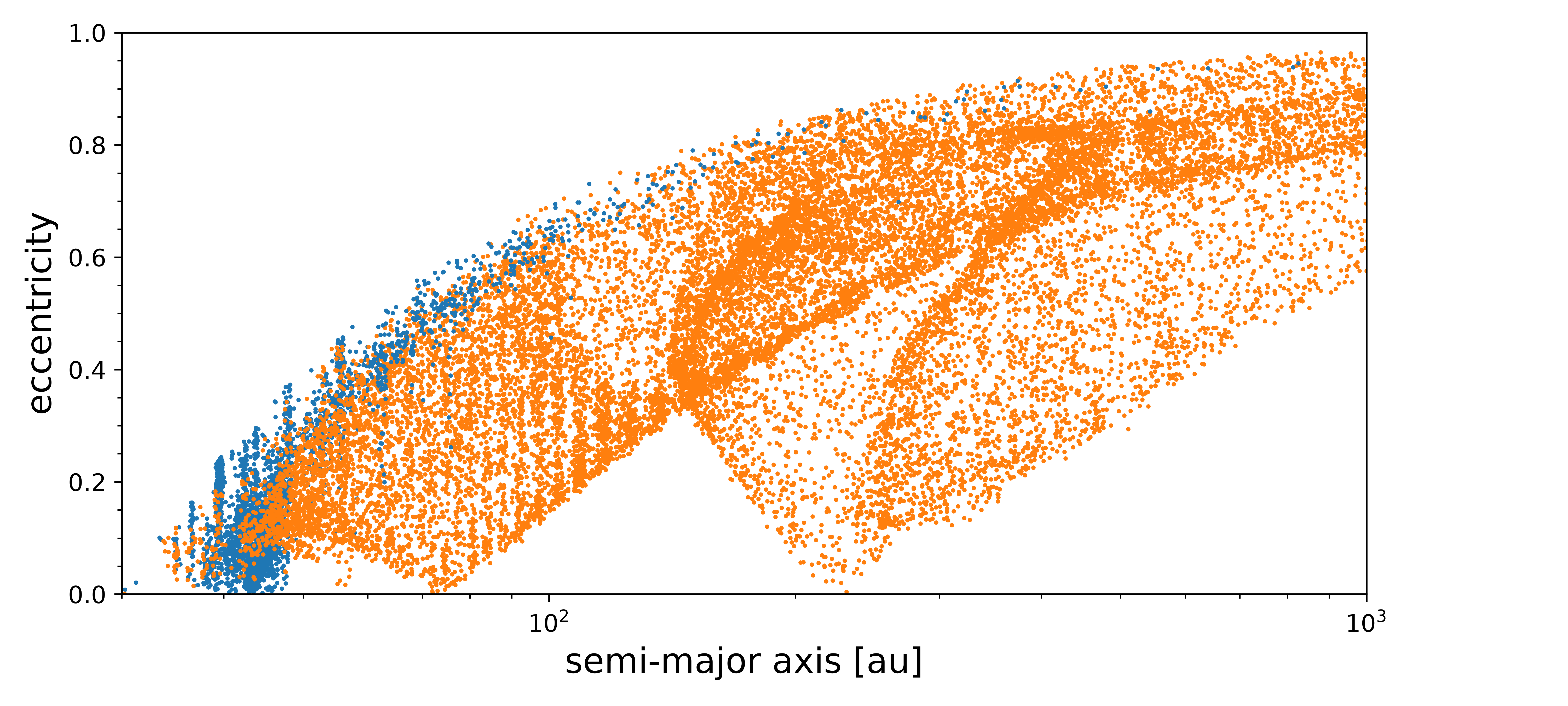}
    \textbf{(d)}
  \end{minipage}

\caption{Comparison of simulation results to the observed TNO population. The panels show (a) the simulation result immediately after the flyby, (b) 4.5~Gyr after the flyby, (c) the TNOs listed in the MPC catalogue (July 2026), and (d) the observed TNOs in blue and simulated particles in orange. The colour in (a)-(c) indicates the inclination. For the observed TNOs, only objects with $q > 30$~au are shown.}
\label{fig:comparison}
\end{figure}

Figure~\ref{fig:comparison} shows (a) the simulation result immediately after the flyby, (b) after 4.5~Gyr of evolution, (c) the observed TNOs listed in the MPC catalogue (July 2026), and (d) the simulation results (orange) overplotted onto the observed population (cyan). For the observed TNOs, only objects with $q > 30$~au are shown. In \citet{Pfalzner:2024a}, we showed that just after the specified flyby, surprisingly many of the fundamental characteristics of the observed TNO population are already successfully reproduced (compare Fig.~\ref{fig:comparison}a and c). In particular, it generates distinct hot and cold Kuiper belt components, as well as detached, extreme, Sedna-like, and retrograde TNOs. In addition, it provides an explanation for the preferentially retrograde orbits of the irregular moons of Jupiter and Saturn \citep{Pfalzner:2024b}, assuming injected TNOs captured by the planets. However, certain features are not fully recovered just after the flyby, such as an overrepresentation of objects with perihelion distances in the range 30~au~$< q <$~35~au and the absence of resonant populations (compare Fig.~\ref{fig:comparison}a and c).

Panels (b) and (d) in Fig.~\ref{fig:comparison} demonstrate that most of the remaining discrepancies are substantially mitigated when one takes the dynamical evolution over the consecutive 4.5~Gyr into account. Most prominently, overrepresentation in the range 30~au~$< q <$~35~au no longer exists and resonant populations appear. Nearly all objects with perihelion distances 30~au~$< q <$~35~au are ejected or relocated to different orbital parameters. The resonant population emerges as a result of gravitational interactions with Neptune. As mentioned above, the detailed properties and evolution of these resonant populations are addressed in a separate paper.

Now the curvature in the distribution is also very similar (see Fig.~\ref{fig:comparison}d). The possible origin of the remaining differences we discuss in Sec.~\ref{sec:discussion}.

\section{TNO colours}
\label{sec:colours}

In addition to replicating the dynamical characteristics, models of the TNO population must accurately reproduce the observed colour distribution. The TNOs show a complex red-to-gray colour distribution, with the cold KBO population being dominated by very red objects, while TNOs at inclinations $i >$~21° \citep{Marsset:2019} and eccentricities $e >$~0.42 \citep{Ali:2021} show a scarcity of very red objects. \citet{Pfalzner:2025} showed that the same flyby that can account for the TNO dynamics would also explain the correlation between their colours and orbital characteristics. The combined explanation of these TNO properties strengthens the evidence for a close flyby of another star to the young Solar System.

Nevertheless, this analysis did not encompass the full 4.5~Gyr timescale of Solar System evolution. Therefore, a critical question remains: does the underabundance of very red TNOs in these regions of parameter space persist throughout Gyr-scale evolution?

\begin{figure}
\centering
  \begin{minipage}[b]{0.44\textwidth}
    \centering
    \includegraphics[width=\textwidth]{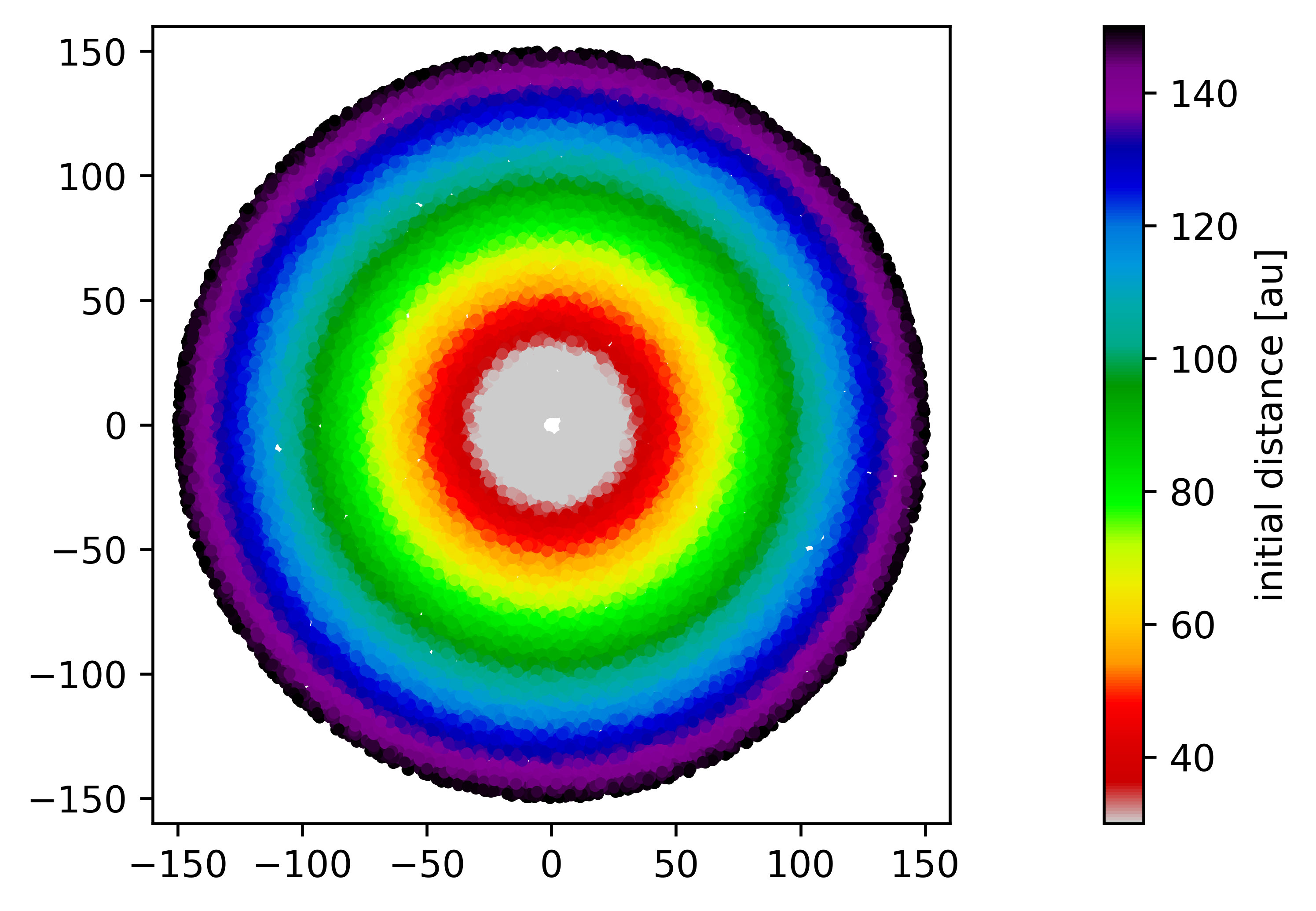}
    \textbf{(a)}
  \end{minipage}

  \centering
  \begin{minipage}[b]{0.44\textwidth}
    \centering
    \includegraphics[width=\textwidth]{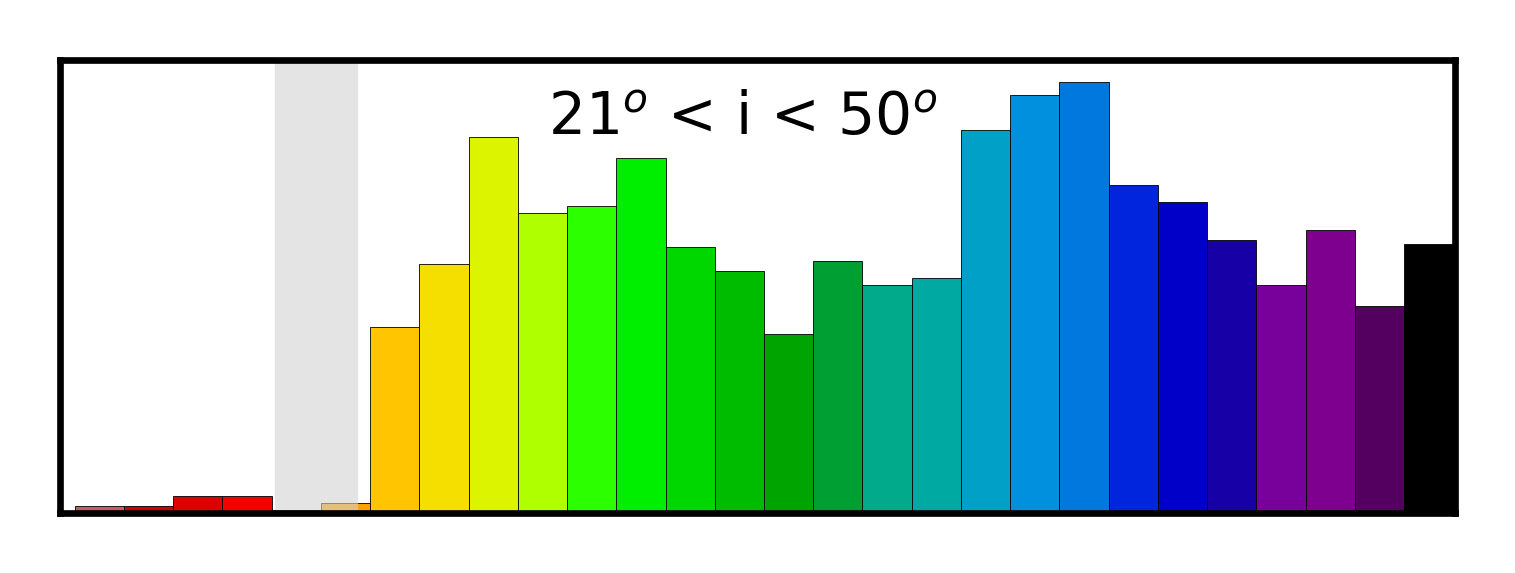}
    \includegraphics[width=\textwidth]{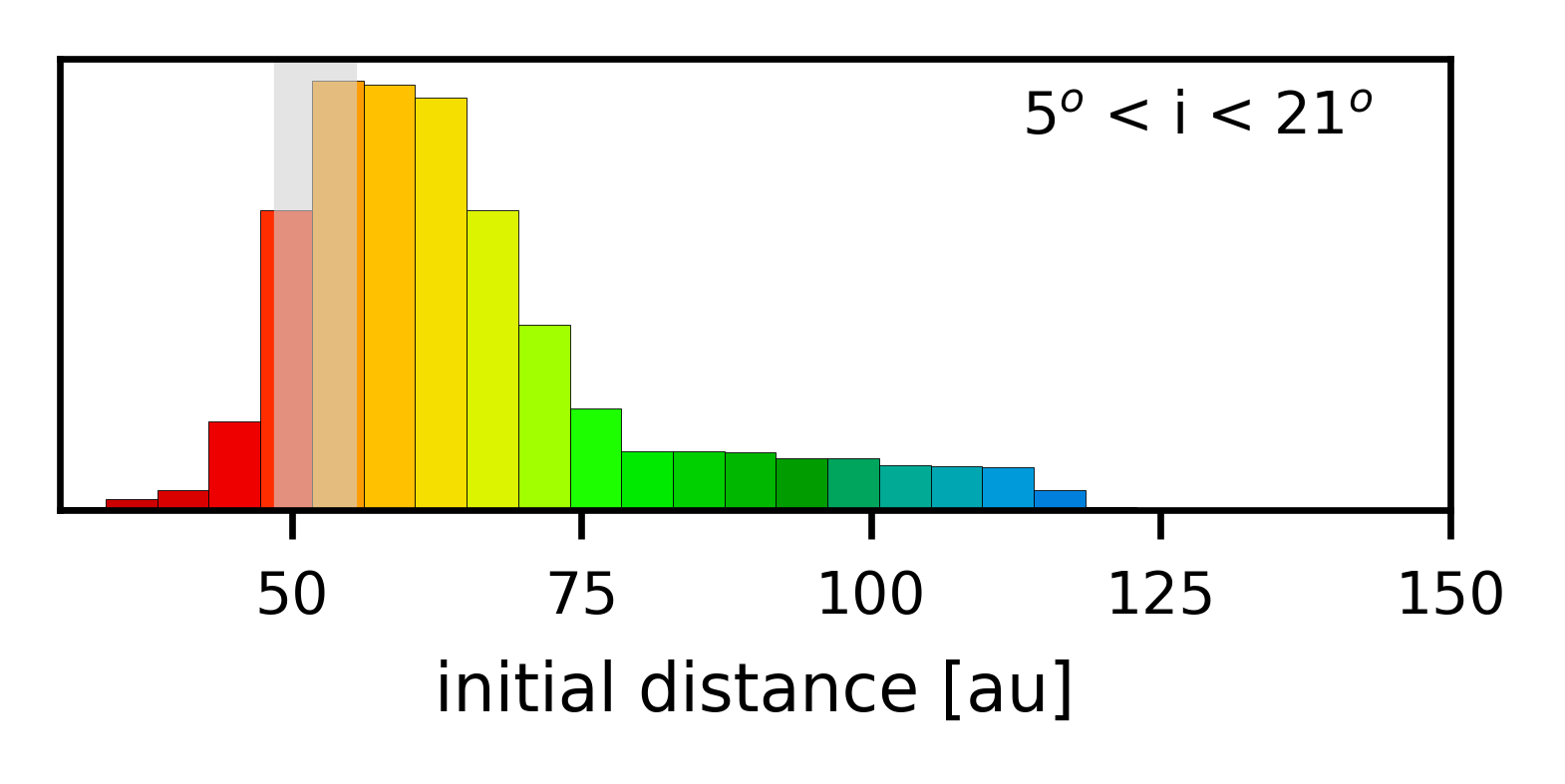}
    \textbf{(b)}
  \end{minipage}

 \centering
  \begin{minipage}[b]{0.44\textwidth}
    \centering
    \includegraphics[width=\textwidth]{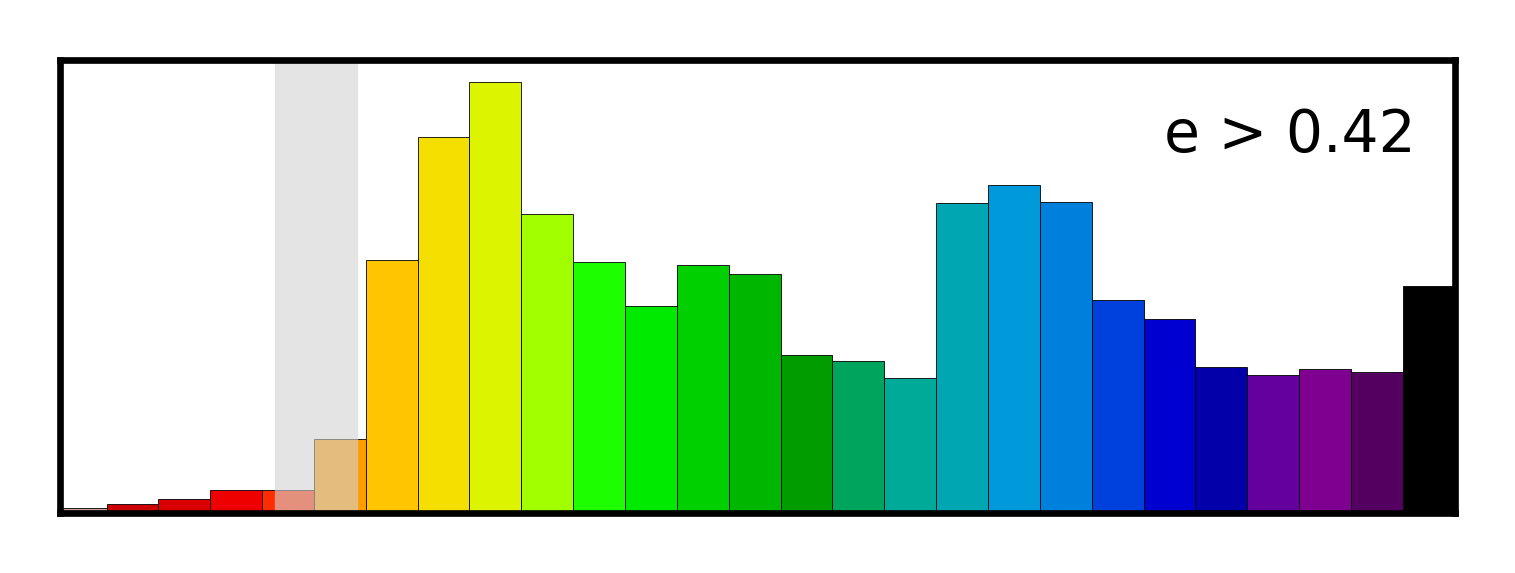}
     \includegraphics[width=\textwidth]{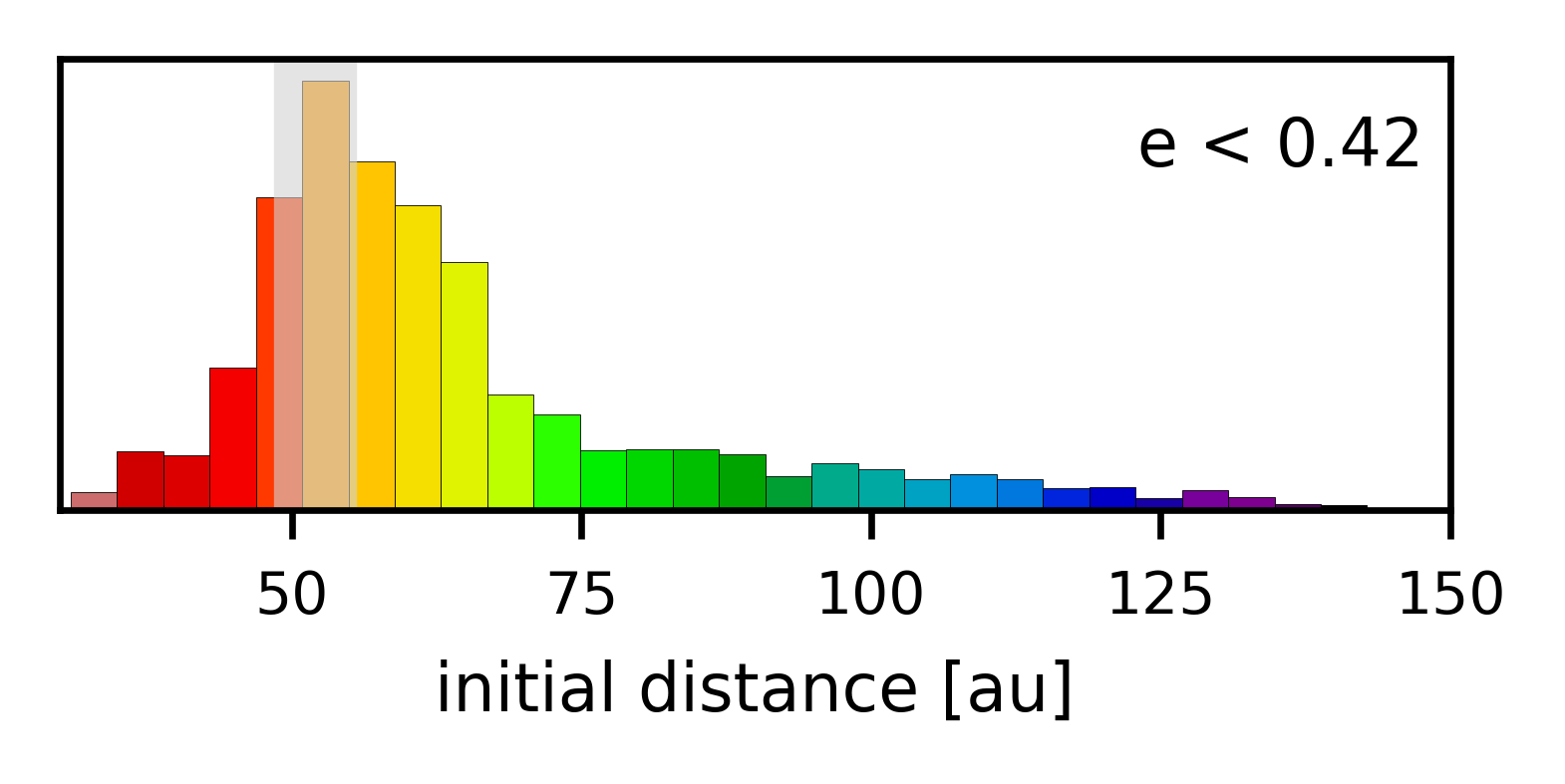}
    \textbf{(c)}
  \end{minipage}

\caption{Evolution of the TNO colour distribution. (a) shows the initial setup, with red TNOs close to Neptune and different colours beyond. Panels (b) and (c) show the situation after the flyby and the consecutive 4.56~Gyr. The observed lack of red TNOs among the high-inclinations and high-eccentricities TNOs is also present after 4.56~Gyr.}
\label{fig:colours}
\end{figure}

Following the approach of \citet{Pfalzner:2025}, we postulate an initial radial colour gradient within the protoplanetary disc. It is widely accepted that cold classical KBOs have undergone minimal radial migration since their formation. Consequently, it is reasonable to infer that the primordial disc was predominantly composed of very red bodies in the cold Kuiper Belt region, while objects located at greater heliocentric distances exhibited a broader diversity of surface colours, ranging from neutral to gray hues. This conceptual framework is represented as a continuous colour gradient spanning 30~au to 150~au, as illustrated in Fig.~\ref{fig:colours}a, serving as a provisional model for the observed spectral slope $S$. Thus, the initial position is the disc stands for a particular colour.

Figure~\ref{fig:colours}b compares the colour distribution of TNOs with inclinations of 5°~$< i <$~21° and 21°~$< i <$~50° after 4.56~Gyr. The scarcity of very red TNOs at high inclinations is retained over this long timespan. Similarly, the eccentricity distribution for $e >$~0.42 still shows a lack of very red TNOs compared to those with $e <$~0.42 (see Fig.~\ref{fig:colours}c). Thus, the similarity between the observed and modelled colour distribution remains even after 4.5 Gyr of evolution.

\section{Discussion}
\label{sec:discussion}

\subsection{Model limitations}

Although this flyby scenario effectively replicates many observed characteristics, a limited number of discrepancies remain between empirical observations and our simulation outcomes, which warrants further scrutiny. Such deviations are expected, as the flyby parameters were optimised to reproduce the system immediately after the event and to align with the present-day TNO population. Consequently, further refinement of these parameters is anticipated as our understanding of the system's long-term dynamical evolution advances \citep{Pfalzner:2024a}.

Nevertheless, the comparison between observational and simulation results is intrinsically constrained by observational biases, as current surveys predominantly detect large, nearby TNOs with low inclinations and eccentricities. Simulations, on the contrary, necessarily incorporate simplifying assumptions. Thus, residual discrepancies must be interpreted within the context of these methodological limitations. The simulations differ in: (1) a substantial fraction of objects exhibiting large perihelion distances; (2) the mean semi-major axis of the cold classical population being found at 38~au, whereas observational data place this population between 42~au and 45~au; (3) there is a diminished proportion of cold KBOs with eccentricities $e <$~0.1; and (4) $\approx$~75\% of Sedna-like objects display inclinations near 11° to 12°, while the simulations forecast a predominance of Sedna-like objects at higher inclinations.

The first point could stem from observations being limited to relatively nearby objects. Thus, our model predicts the existence of such distant objects and constitutes a test of our flyby model. In recent years, more distant TNOs have been discovered, for example by the New Horizons mission \citep{Fraser:2024}, which could indicate that such a population still awaits discovery. 

Points (2) and (3) are likely connected to a significant limitation of the present model. Due to computational costs, we excluded mutual gravitational interactions among TNOs and considered only the effect of the giant planets on the massless test particles. \citet{Punzo:2014} examined the evolution of the KBO population following a distinct stellar flyby ($M_p =$~1~\MSun, $a_p =$~200~au, $i_p =$~90°), simulating the first 0.5~Myr post-encounter with explicit inclusion of TNO–TNO interactions between the larger objects. Their results indicate that secular evolution facilitates partial repopulation of the low-eccentricity component. In addition, \citet{Munoz:2025} assessed the influence of the most massive TNOs -- most notably Pluto -- on the stability of resonant populations. They demonstrate that these massive bodies significantly alter decay rates, not only for the 3:2 resonance but also for the 2:1 resonance. Therefore, incorporating mutual interactions among TNOs will likely reproduce the observed cold classical KBO population more accurately.

Furthermore, our simulations employ a constant distribution of test particles to ensure adequate spatial resolution in the outer regions of the disc. This methodology also permits post-processing application of various surface density profiles to the simulation results. However, to accurately incorporate inter-TNO gravitational forces, it will be necessary to transition from a test particle framework to one representing a realistic mass distribution within the TNO population.

\bigskip
\bigskip

\subsection{Comparison to other models}

A leading model for the dynamics of the KBO population is the Nice model and its descendants \citep{Fernandez:1984,Gomez:2003,Gladman:2021}. Recently, \citet{Lykawka:2026} conducted a comprehensive study that is particularly suitable for direct comparison with the flyby scenario evaluated here. Employing \mbox{$N$-body} integrations that span 4.5~Gyr, their simulations adopted canonical Nice model initial conditions: a massive primordial planetesimal disc, the four giant planets, and 1500 Pluto-mass perturbers to facilitate Neptune's outward migration. Their principal results include: (a) successful reproduction of the principal TNO sub-populations -- namely cold and hot classicals, resonant, scattered, and detached objects; (b) robust resonant capture preferentially occurring at low-order Neptunian mean motion resonances (MMRs; e.g., 3:2, 2:1); and (c) optimal agreement with empirical data arising in models incorporating a ``jumping'' Neptune, mutual gravitational interactions among Pluto-class bodies, and a primordial disc truncated at 45~au to 47~au. Persistent difficulties were also noted: (1) underproduction of TNOs in distant resonances (beyond 50~au); (2) introduction of a primordial scattered disc augmented distant MMR occupation but resulted in an overabundance of scattered disc objects; (3) all tested configurations failed to efficiently generate detached ($q >$~40~au) and very high-inclination $(i >$~45°) TNOs; (4) no scenario reproduced the observed abundance of extreme populations, such as distant ($a >$~245~au), low-inclination detached or extreme TNOs ($q >$~50~au or $i >$~50°); (5) the dynamical effects of stochastic (``grainy'') migration diminished rapidly as Pluto-class bodies were depleted; and (6) to maintain a remnant Pluto population within 50 au, the initial number of Pluto-class objects had to be on the order of about 150 -- 500, consistent with the current census. In conclusion, while such models can replicate much of the structure interior to 50 au, they remain deficient in accounting for the formation of detached, high-inclination, or extreme TNOs. \citet{Lykawka:2026} therefore propose that additional dynamical mechanisms -- such as perturbations from an as-yet undetected distant planet -- may be required to fully elucidate the architecture of the outer Solar System. 

Currently, it is unclear whether the Nice models also produced the observed TNO colour distribution. In particular, it would be interesting to see whether this distribution match can be obtained after a 4.5 Gyr evolution.

In summary, both models reproduce the main dynamical TNO populations -- namely cold and hot classicals, resonant, scattered, and detached objects. However, they differ in their strengths and weaknesses. Specific versions of the Nice model provide better matches for the cold population, but exhibit substantial difficulties in replicating the more distant populations, including those in distant resonances, detached orbits, and objects with high-inclination or retrograde trajectories. By contrast, the stellar flyby robustly generates the observed distant and high-inclination populations, although it requires parameter refinement to accurately reproduce the spatial distribution of the cold classicals. Another strength of the stellar flyby is that the same flyby that fits the dynamics can also account for the colour distribution. Such refinements are plausibly achievable via modest adjustments to the flyby model parameters. Future observations -- particularly those targeting more distant and dynamically extreme TNOs -- will provide critical constraints on both theoretical frameworks, facilitating a more definitive assessment of their relative validity or potentially indicating the need for a hybrid paradigm to fully account for the outer Solar System's dynamical architecture.

\section{Summary and Conclusion}

We modelled the dynamical evolution of the TNO population for 4.56~Gyr after a stellar flyby. As starting conditions, we chose the flyby that provides a good fit to today's TNO distribution immediately after the flyby. The question was how the simulated population's dynamics would change over extended timescales through the interactions with the Solar System's planets.

Our results show that this agreement is not only maintained, but actually improves significantly. The most notable improvements include that
\begin{itemize}
\item TNOs with 30~au~$< q <$~35~au are ejected, so the lack of objects in this range is now recovered,
\item the relationship between eccentricity and semi-major axis is even better recovered, and
\item the TNOs in resonant orbits with Neptune develop, including the most populated resonances (2:1, 3:2, 5:2).
\end{itemize}

Detached, Sedna-like, and retrograde TNOs largely retain their original dynamical parameters. This stability is reassuring, as it confirms their robustness and supports their use as benchmarks for assessing the quality of the fit immediately after the flyby. In addition to the dynamics, the colour distribution of TNOs is maintained over 4.56~Gyr. Because the dynamic and colour constraints are simultaneously fulfilled, this considerably strengthens the argument for such a close flyby.

In our study, we considered only the effects of planetary perturbations on TNO dynamics. Future work should investigate how mutual interactions within the TNO population influence their long-term evolution. Only after accounting for these effects will it be meaningful to fine-tune the parameters of the Solar System flyby, as the cold population still requires further adjustment.

\begin{acknowledgments}
SP acknowledges funding in the context of the Cluster of Excellence ``Our Dynamic Universe'' (Dynaverse) supported by the DFG fund EXC 3707 with the project number 533607693 and the Excellence Network DynaverseJSC funded by the Helmholtz Association. The research data management was carried out within the framework of the PUNCH4NFDI consortium supported by DFG fund NFDI 39/1, Germany, and it fulfils the requirements listed in \citet{Pfalzner_FAIR}.
\end{acknowledgments}
\newpage
\appendix
\setcounter{figure}{0}    
\renewcommand\thefigure{A.\arabic{figure}}    
\noindent
Figure \ref{fig:Families_sdr} shows the temporal evolution of the dynamic families that are least affected by the interactions with the giant planets -- namely, the Sedna-like objects, the detached TNOs and the retrograde TNOs.

\begin{figure*}[h]
\centering
    \begin{minipage}[b]{0.30\textwidth}
    \centering
    \includegraphics[width=\textwidth]{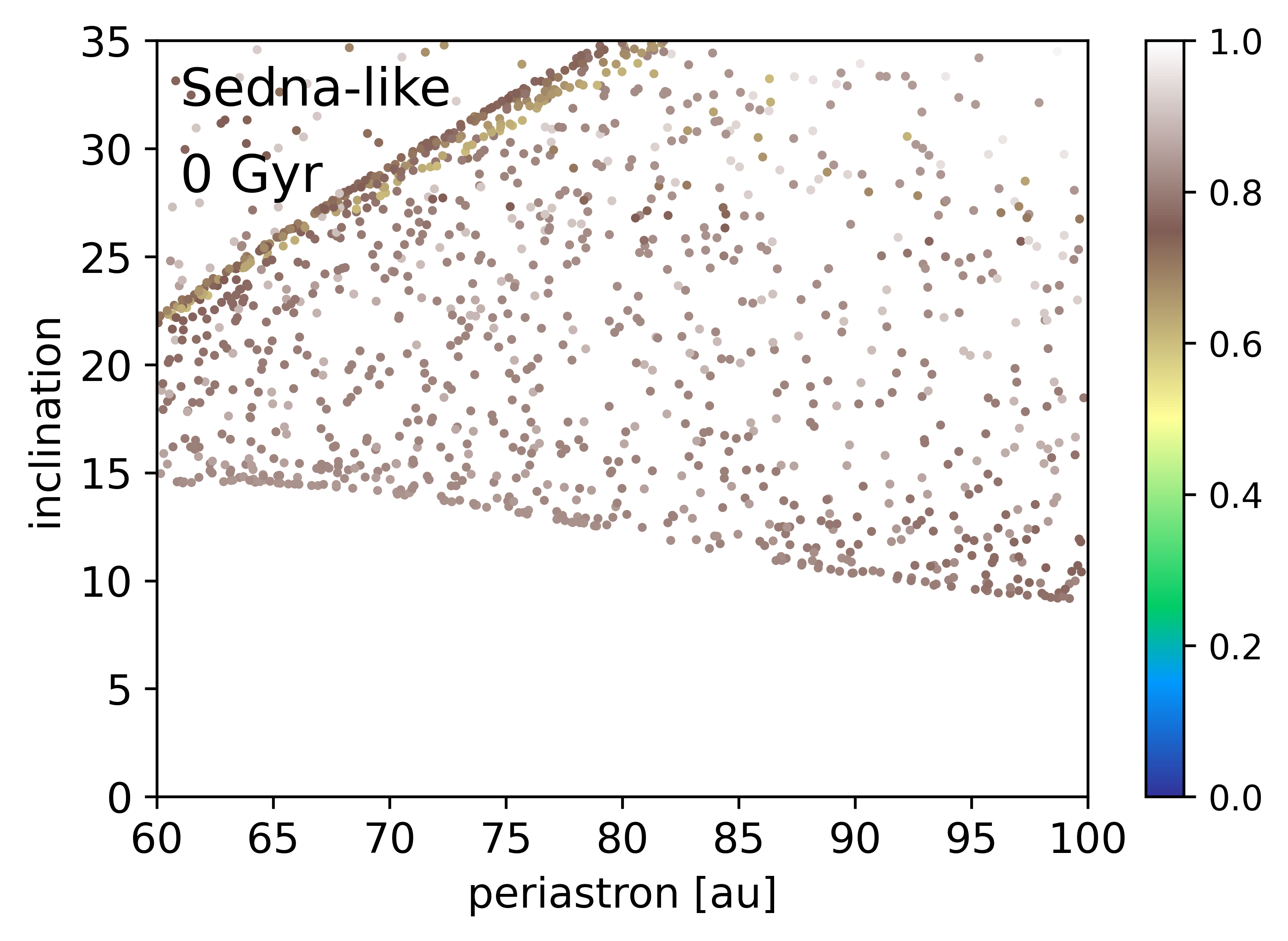}
  \end{minipage}
\centering
    \begin{minipage}[b]{0.30\textwidth}
    \centering
    \includegraphics[width=\textwidth]{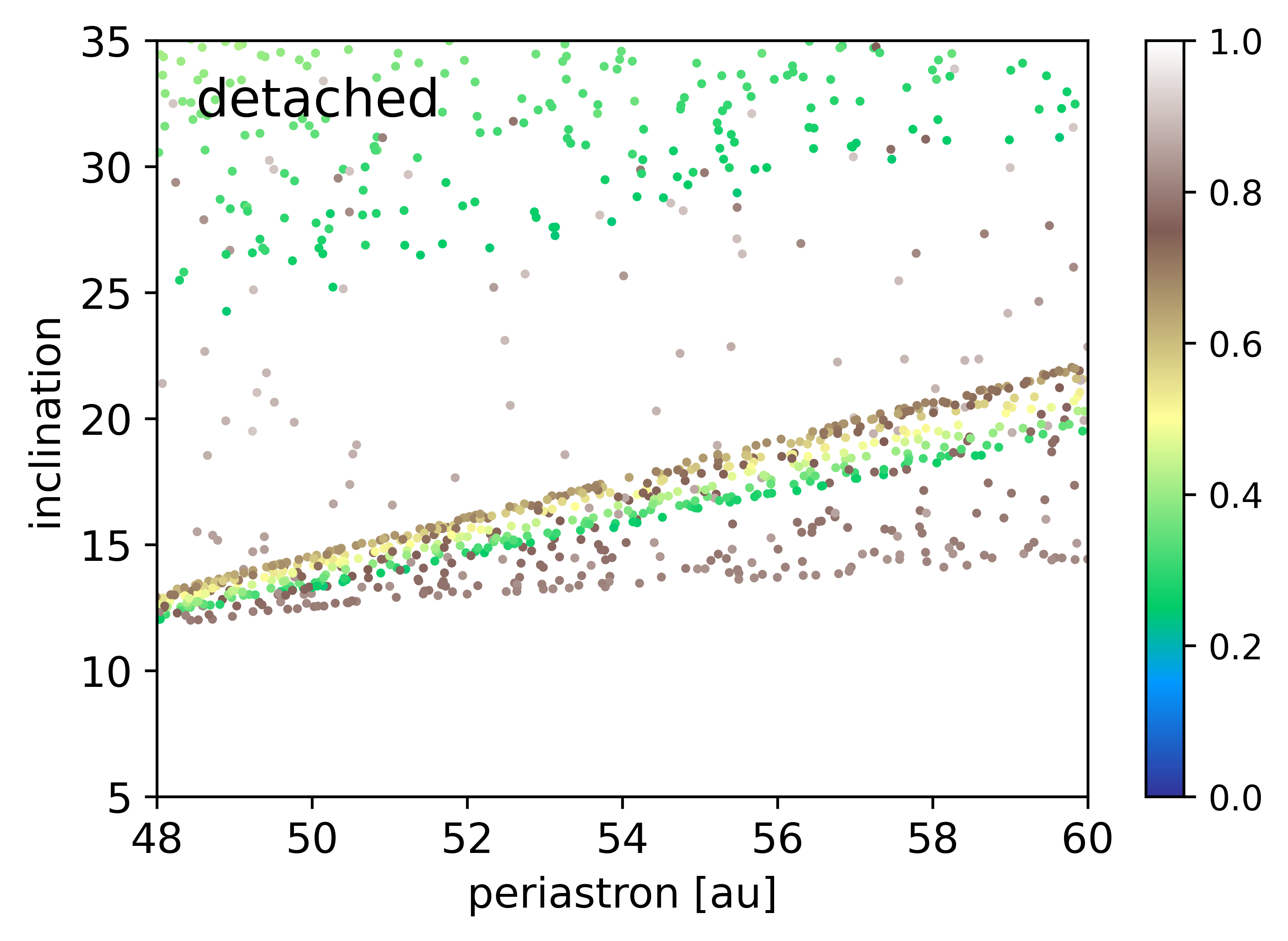}
  \end{minipage}
\centering
   \begin{minipage}[b]{0.30\textwidth}
   \centering
    \includegraphics[width=\textwidth]{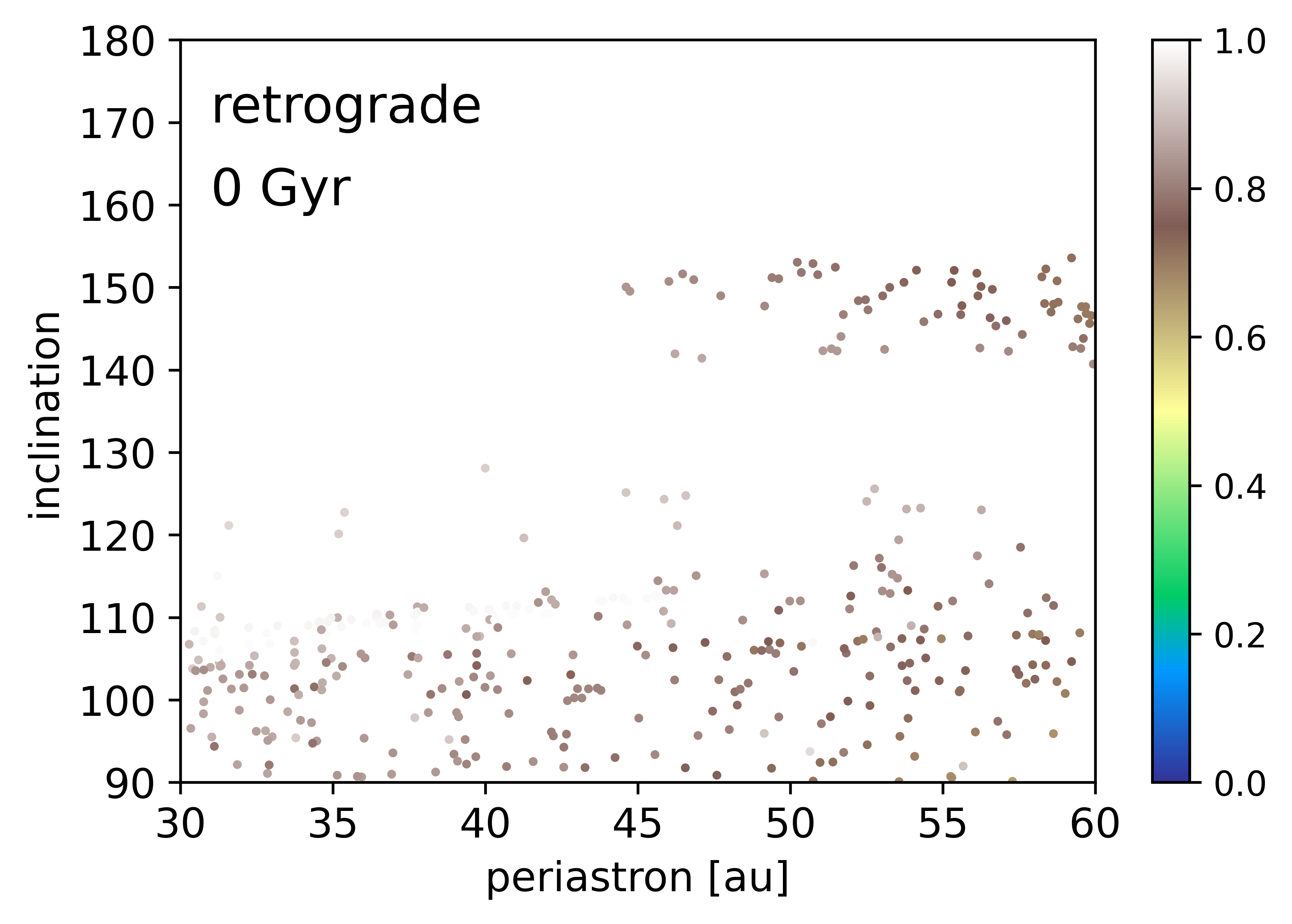}
  \end{minipage}
\centering
    \begin{minipage}[b]{0.30\textwidth}
    \centering
    \includegraphics[width=\textwidth]{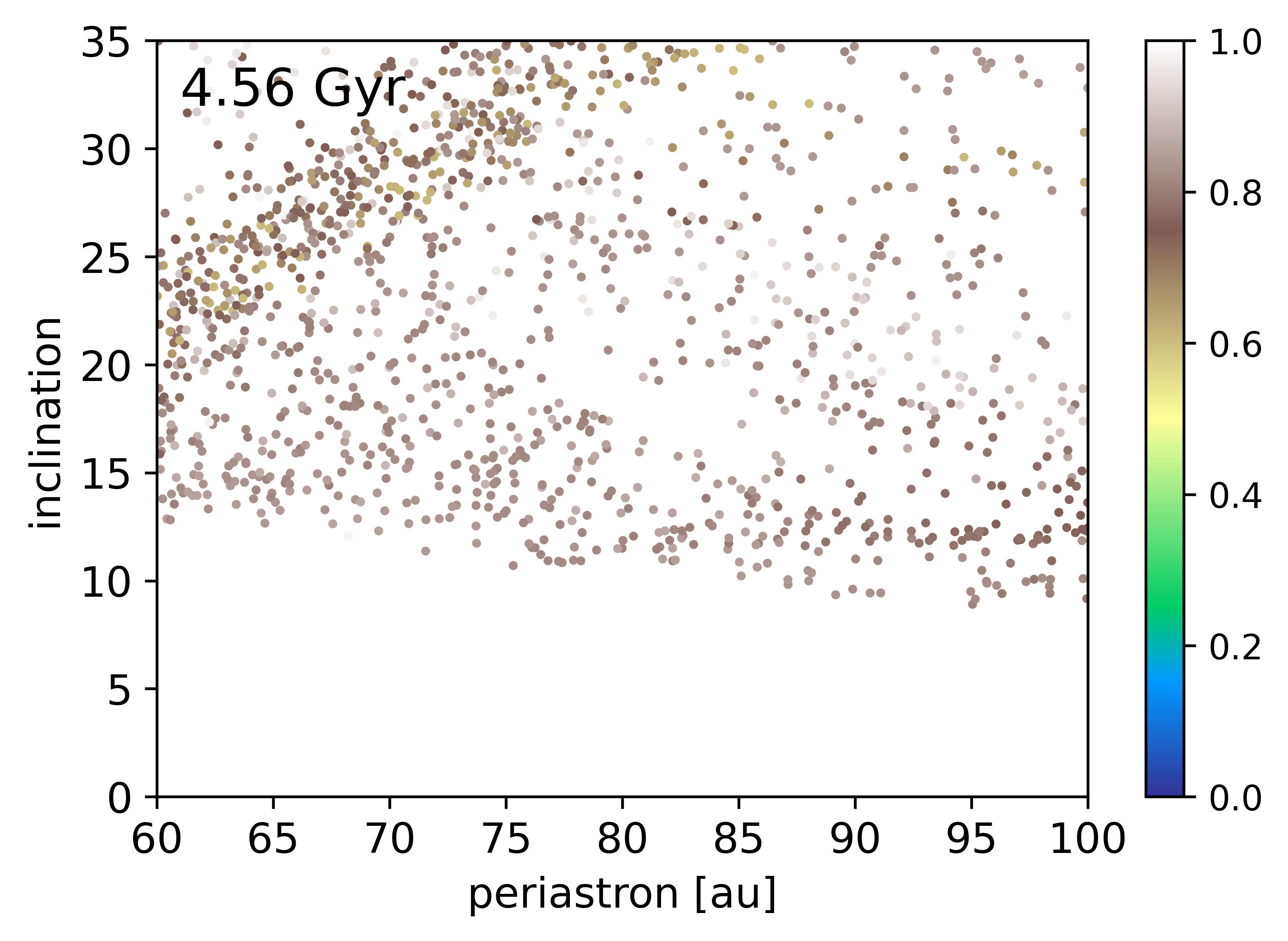}
    \textbf{(a)}
  \end{minipage}
\centering
    \begin{minipage}[b]{0.30\textwidth}
    \centering
    \includegraphics[width=\textwidth]{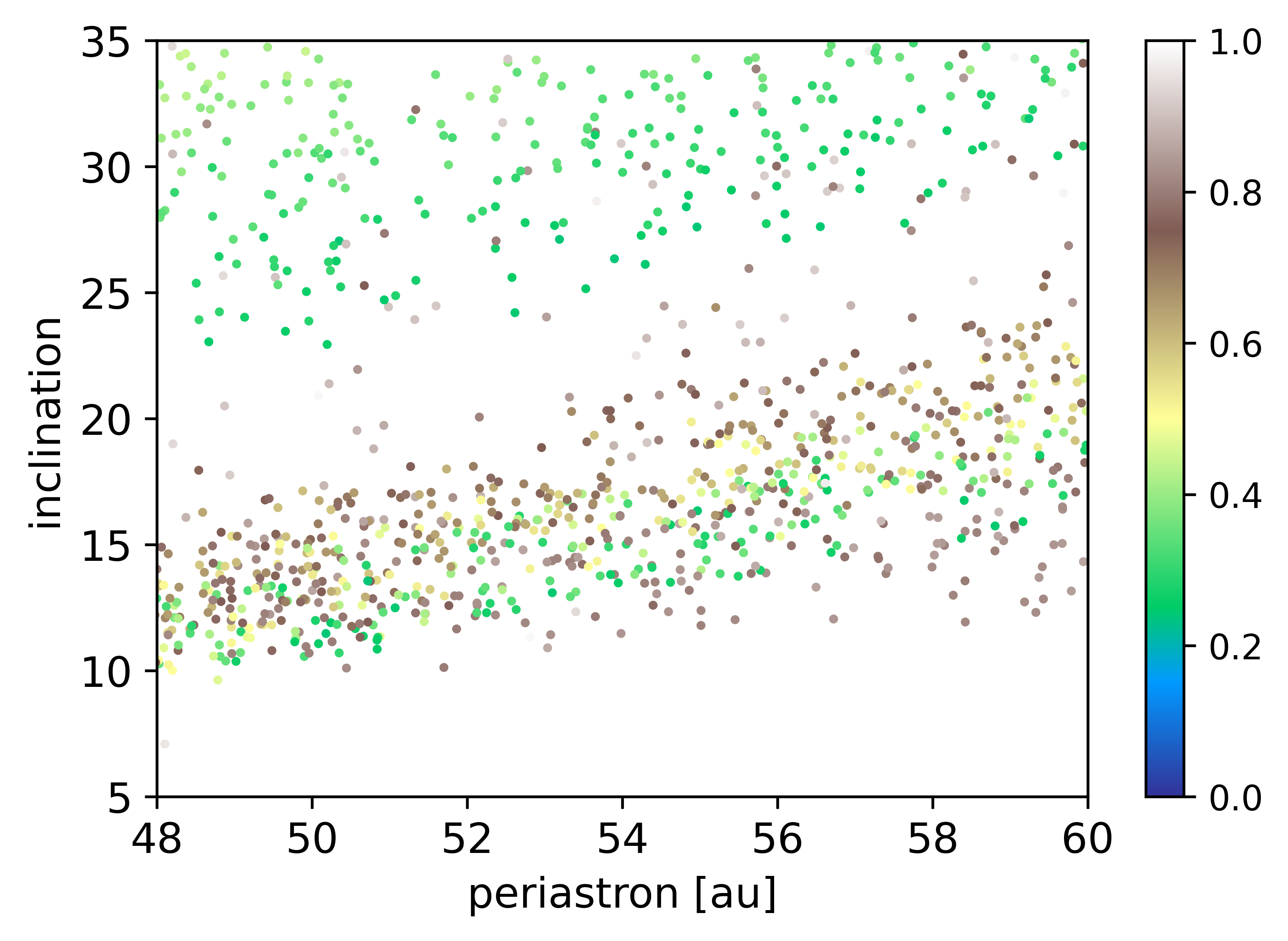}
    \textbf{(b)}
  \end{minipage}
\centering
    \begin{minipage}[b]{0.30\textwidth}
    \centering
    \includegraphics[width=\textwidth]{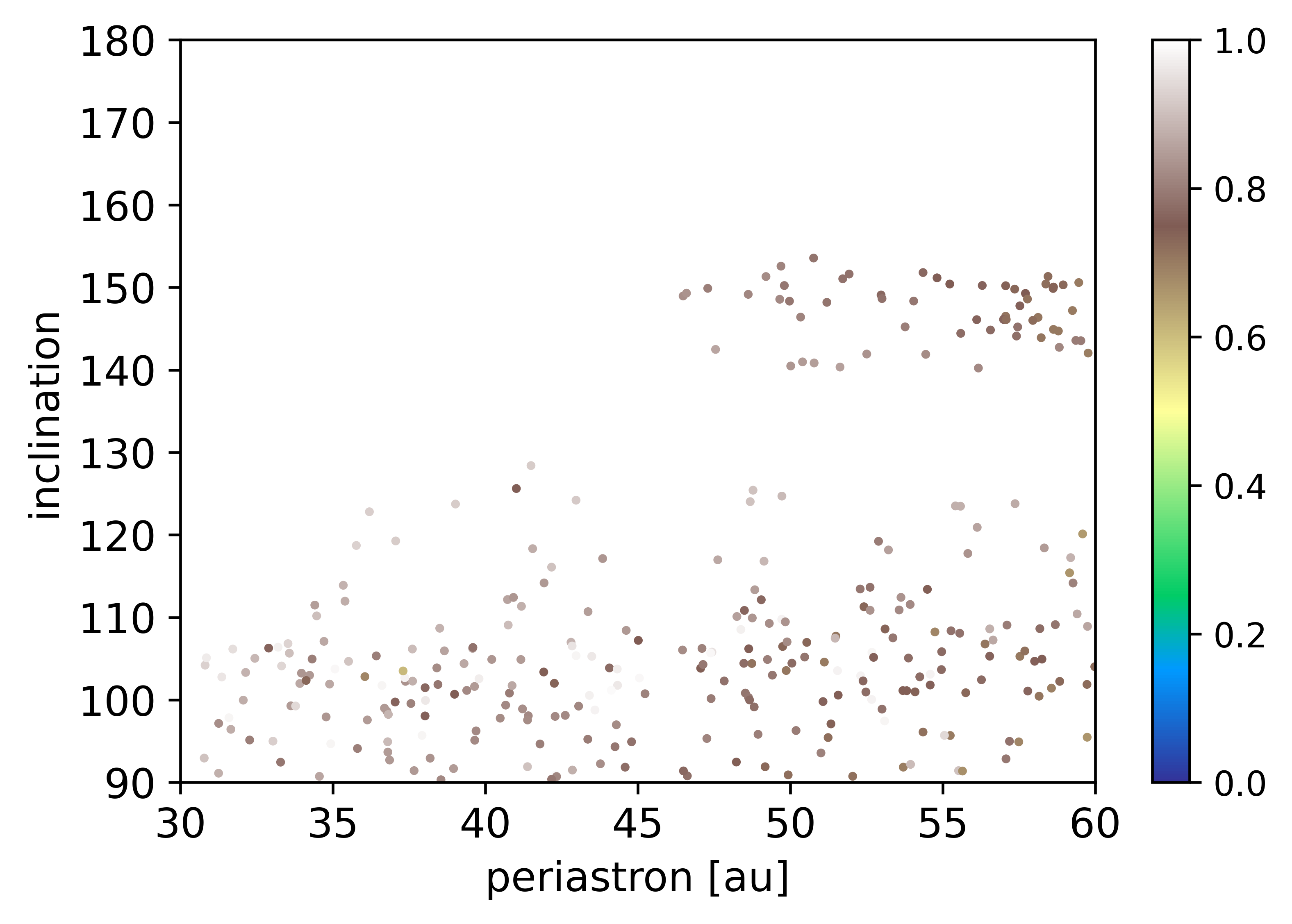}
    \textbf{(c)}
  \end{minipage}
\caption{Evolution of Sedna-like, detached and retrograde TNOs' dynamics. Shown are the inclination vs the periastron distance of the test particles in our simulations. The colour indicates their eccentricity. The top row shows the dynamic properties directly after the flyby, the bottom row the same after 4.56~Gyr of dynamic evolution for a) Sedna-like, b) detached and c) retrograde TNOs.}
\label{fig:Families_sdr}
\end{figure*}

\newpage

\bibliography{references}{}
\bibliographystyle{aasjournal}

\end{document}